# Title

Miniaturized optically-generated Bessel beam ultrasound for volumetric transcranial brain stimulation

# Authors

Yueming Li[1], Guo Chen[1], Tiago R. Oliveira[2,6], Nick Todd[2], Yong-Zhi Zhang[2], Carolyn Marar[3], Nan Zheng[4], Lu Lan[1], Nathan McDannold[2], Ji-Xin Cheng[1,3,*], Chen Yang[1,5,*]

# Affiliations

[1] Department of Electrical and Computer Engineering, Boston University, Boston, MA 02215, USA.

[2] Department of Radiology, Brigham and Women's Hospital, Harvard Medical School, Boston, MA, United States.

[3] Department of Biomedical Engineering, Boston University, Boston, MA 02215, USA.

[4] Division of Materials Science and Engineering, Boston University, Boston, MA 02215, USA.

[5] Department of Chemistry, Boston University, Boston, MA 02215, USA.

[6] CECS - Center for Engineering, Modeling and Applied Social Sciences, Federal University of ABC (UFABC), São Bernardo do Campo, SP 09606405, Brazil.

* Corresponding author. Email: cheyang@bu.edu (C.Y.); jxcheng@bu.edu (JX.C.).

# Abstract

Non-invasive stimulation of small, variably shaped brain sub-regions is crucial for advancing our understanding of brain functions. Current ultrasound neuromodulation faces two significant trade-offs when targeting brain sub-regions: miniaturization versus volumetric control and spatial resolution versus transcranial capability. Here, we present an optically-generated Bessel beam ultrasound (OBUS) device designed to overcome these limitations. This 2.33 mm-diameter miniaturized device delivers a column-shaped field achieving a lateral resolution of 152 μm and an axial resolution of 1.93 mm, targeting brain sub-regions with an elongated volume of tissue activation. Immunofluorescence imaging of mouse brain slices confirms its ability to stimulate cells at a depth of 2.2 mm. Additionally, OBUS outperforms conventional Gaussian ultrasound in transcranial transmission efficiency and beam shape preservation. Electrophysiological recordings and functional MRI captured rodent brain responses evoked by OBUS, demonstrating OBUS's ability to non-invasively activate neural circuits in intact brains. This technology offers new possibilities for studying brain functions with precision and volumetric control.

**MAIN TEXT**

**Introduction**

Precisely modulating brain subregions is critical for decoding brain function at the level of individual functional units, enabling researchers to explore complex neural mechanisms and developing more effective therapeutic interventions (*1–3*). An ideal approach to this task requires key features such as non-invasiveness, miniaturization, and advanced precision. Specifically, it should achieve spatial resolution on the order of a few hundred micrometers and be adaptable to the complex and heterogeneous morphologies of brain subregions (*4*, *5*). For example, ocular dominance columns (ODCs) in humans exhibit an elongated structure spanning multiple cortical layers over 2 to 3 mm, with a mean width of only 863 µm (*5*). In macaque monkeys, ODCs are even narrower, with widths ranging from 400 to 700 µm, posing significant challenges for precise targeting (*6*, *7*). Additionally, the interleaved organization of ODCs corresponding to the left and right eyes introduces further complexity, as off-target stimulation could disrupt decoding of these functional subregions (*6*). These intricate anatomical and functional characteristics highlight the critical need for stimulation tools capable of precise volumetric control to accurately target specific brain subregions while minimizing the risk of activating adjacent structures.

Current neuromodulation tools have made achieved precise control over the stimulation volume. Deep brain stimulation (DBS), the gold standard in neuromodulation for decades, is widely used in the clinical treatment of Parkinson's disease and tremors (*8–10*). DBS employs current steering techniques with directional leads comprising radially segmented electrodes, offering superior targeting of non-spherical brain regions and precise control over the volume of tissue activation (VTA) to improve therapeutic outcomes (*11*, *12*). To mitigate DBS vulnerability to heterogeneous tissue properties like impedance mismatch, computational models evaluate electrode configurations (*13*) and multi-set steering effects (*14*). Despite these advancements, DBS remains invasive, and its VTA in human brains is still on the millimeter-scale. Optogenetics has emerged as a powerful neuromodulation technology (*15–17*), enabling cell-type-specific modulation of neural circuits, including the selective activation of individual ODCs in non-human primates (*18*). However, its broader application faces several limitations: (1) Accessibility is largely limited by the complexity of the animal model, viral vector design and handling, and high costs (2) effective implementation requires long incubation periods (~3 months) following viral vector injection (*19*); (3) limited light penetration into brain tissue results in ~99% absorption at a depth of 0.9 mm, generating heat and posing risks of thermal side effects (*17*, *19*); (4) incomplete stimulation of elongated structures such as ODCs, which span 2–3 mm. Although depth-specific, multi-site optical probes have been developed to target deeper brain regions (*20*), invasive fiber-based delivery of photons is needed.

Non-invasive neuromodulation techniques such as transcranial direct current stimulation (tDCS) (*21*) and transcranial magnetic stimulation (TMS) (*22*) suffer from poor spatial resolution, ranging from several millimeters to centimeters. Transcranial focused ultrasound (tFUS), in contrast, offers superior resolution on the order of a few millimeters at frequencies below 1 MHz, enabling efficient transcranial penetration (*23*, *24*). Its functional efficacy has been demonstrated on mice, monkeys, and humans(*25–29*). tFUS also allows control over the VTA through acoustic lenses (*30*) or 2D arrays (*31*, *32*). However, acoustic lenses introduce additional interfaces that compromise acoustic energy delivery, while 2D arrays require complex engineering for precise multichannel control over timing and amplitude to shape wavefronts. Moreover, the bulkiness and rigidity of current ultrasound arrays limit their portability, limiting their use as wearable devices for awake animal studies or clinical applications (*33*).

Optoacoustic neuromodulation is an emerging technique that offers a promising alternative for ultrasound neural stimulation, providing enhanced spatial resolution and greater flexibility in waveform engineering (*34*). This technique leverages pulsed laser-induced ultrasound waves by utilizing a combination of light-absorbing materials and a thermally expandable medium. When coated onto optical fibers, this composite forms a point ultrasound source capable of achieving sub-cellular resolution (*35–37*). Furthermore, the use of soft materials allows for the fabrication of high-numerical-aperture curvatures, mitigating the cracking risk due to the mechanical rigidity of conventional piezoelectric transducers. This advantage surpasses the performance of traditional acoustic lenses by enabling spatial resolutions closer to the theoretical diffraction limit at a given frequency, without requiring higher frequencies that would compromise transcranial penetration (*38*).

Here, we introduce an optically-generated Bessel-beam ultrasound (OBUS) device - a miniaturized, non-invasive, and non-genetic neuromodulation tool. OBUS incorporates two key innovations: 1. The use of a Bessel beam ultrasound enables OBUS to achieve volumetric targeting of column-shaped brain subregions, such as ODCs, and offers superior transcranial ability compared to conventional Gaussian beam. 2. The use of optically-generated ultrasound enables significant miniaturization, resulting in a compact device with a footprint of 2.33 mm in diameter and a weight of just 2.1 mg. The device is fabricated by embedding candle soot nanoparticles in polydimethylsiloxane (PDMS) at an 8:1 polymer-to-curing agent ratio, optimized for a high optoacoustic conversion efficiency. Upon nanosecond laser illumination, OBUS produces an elongated acoustic field with a lateral resolution of 152 μm and an axial resolution of 1.93 mm, enabling precise volumetric stimulation. Simulations using rat skull models demonstrate OBUS's ability to maintain beam shape and intensity during transcranial propagation, outperforming conventional Gaussian beams. Immunofluorescence validates OBUS's capability to stimulate cortical layers up to 2.2 mm depth in the mouse brain. Electrophysiological recording and functional MRI recorded robust neural and hemodynamic responses elicited by OBUS in rodent brains, showcasing its capacity to modulate neural circuits in intact brains. Safety evaluations confirm compliance with mechanical and thermal thresholds in vivo. OBUS represents a significant tool advancement for non-invasively studying brain function and neural circuits, offering new opportunities to explore complex neural circuits and contribute to future neuromodulation research and clinical applications.

## Results
### Design of OBUS

To effectively target elongated brain subregions such as ODCs, we adopted the Bessel beam concept to optically generate a Bessel beam ultrasound field using an optoacoustic device. This approach is designed to achieve both an elongated volume of tissue activation (VTA) and high transcranial efficiency simultaneously for noninvasive brain stimulation (**Fig. 1A**). The OBUS is illuminated by an optical fiber aligned through a 3D-printed adapter ensuring uniform laser illumination. When positioned on the skull, OBUS generates an elongated ultrasound field that penetrates the skull and reaches the targeted brain region.

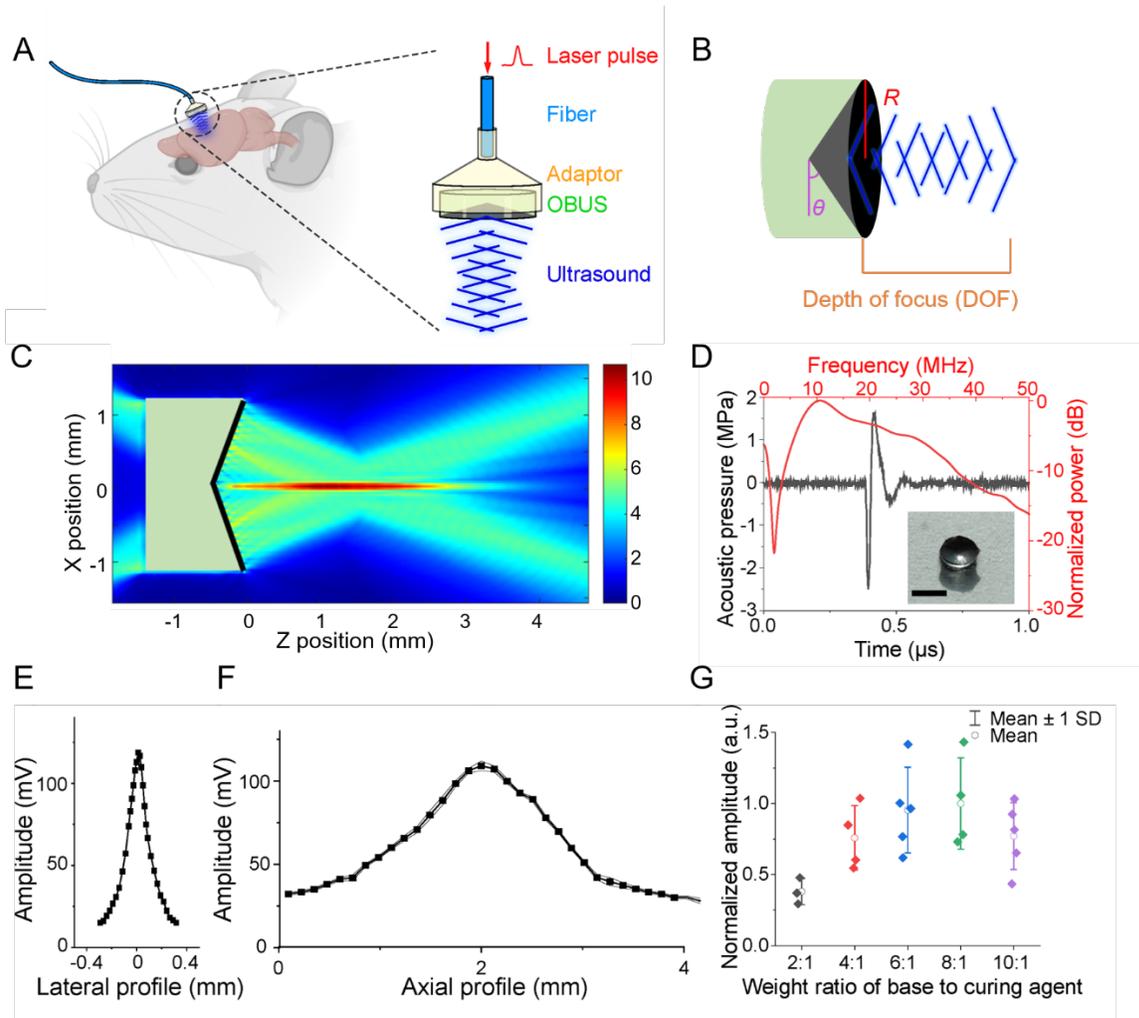

**Fig. 1. Design of OBUS and characterization of the acoustic field generated.**
(A) Schematic illustration of the OBUS device for non-invasive brain stimulation in rodents. (B) Schematic of OBUS device and the resulting acoustic field. Key parameters are highlighted. (C) Simulated acoustic field generated by OBUS with a conical angle of 15°. (D) Acoustic waveform and corresponding frequency spectrum recorded at the peak pressure location. Inset: Photo of the OBUS device. Scale bar: 2 mm. (E) Lateral profile of the acoustic field generated by OBUS. (F) Axial profile of the acoustic field generated by OBUS. (G) Optoacoustic signal intensity as a function of the PDMS base-to-curing agent weight ratio.

To generate the acoustic Bessel beam, a conical optoacoustic emitter surface, analogous to an axicon lens, was employed to generate a zeroth-order Bessel beam propagating along the axial direction (**Fig. 1B**). Two critical parameters of an elongated acoustic focus used for brain stimulation are the depth of focus (DOF) and the peak pressure position, which corresponds to the VTA length and the region where the highest acoustic energy is delivered, respectively. The DOF can be precisely tuned by adjusting the radius ($R$) and angle ($\theta$) of the conical optoacoustic surface, as defined by the governing equation:

$$\text{DOF} = \frac{R}{\tan(\theta)} \tag{1}$$

To determine the optimal $R$ and $\theta$ values for positioning the peak pressure at the desired VTA depth, we conducted simulations based on four criteria for effective brain stimulation: (1) we decided to demonstrate OBUS in rodents due to resource accessibility, targeting a lateral resolution ≤ 0.35 mm to enable brain mapping in the mouse model (*39*); (2) we ensured that the DOF was shorter than the depth of the mouse brain, which is approximately 6 mm; (3) we aimed to position the maximum point of the field deeper than 1.5 mm beneath the skin to target the cortical layer of the brain; (4) we miniaturized the OBUS to have a diameter of < 4 mm (approximately half the width of a mouse brain) to facilitate ease of manipulation and enable potential multisite stimulation.

To satisfy the outlined criteria, we selected an $R$ value of 1.65 mm. We then simulated the acoustic field generated by OBUS using a 1.65 mm radius with conical angles of 10°, 15°, 18°, 20°, and 30°. The simulations employed a central frequency of 15 MHz central frequency with a 200% bandwidth based on previous work (*38*). Water was chosen as the propagation medium, with material-specific acoustic speed and density values applied accordingly. The simulation results are summarized in **Table 1**. Based on the simulation, a conical angle of 15° and an overall device diameter of 2.33 mm satisfied the criteria for lateral resolution, DOF, and maximum pressure position. The optimal simulated acoustic field using these parameters is shown in **Fig. 1C**.

| Conical angle $\theta$ | Lateral resolution (mm) | Axial resolution (mm) | Maximum pressure position (mm) |
|---|---|---|---|
| 10° | 0.30 | 6.60 | 3.21 |
| 15° | 0.33 | 4.49 | 2.06 |
| 18° | 0.24 | 3.34 | 1.48 |
| 20° | 0.21 | 3.06 | 1.14 |
| 30° | 0.12 | 1.88 | 0.50 |

**Table 1. Simulation results of the acoustic field generated by OBUS with various conical angles. Lateral and axial resolutions are defined by the full width at half maximum of the pressure profile along *x* and *z* directions.**

Our previous work on a soft optoacoustic pad has shown that miniaturization through reducing the diameter while maintaining the same laser density reduces ultrasound intensity at the focus, limiting its potential for further miniaturization (*38*). In contrast, through simulation we confirmed that the diameter of OBUS primarily affects the DOF but not the maximum intensity along the zero-order beam. Simulations confirm that OBUS with diameters from 12.2 mm to 2.33 mm deliver consistent peak intensities, as shown in **Supplementary Fig. S1**. OBUS diameter designed according to desired DOF can output consistent maximum acoustic intensity, making OBUS well-suited for miniaturized transcranial neuromodulation applications.

**Characterization and optimization of OBUS optoacoustic efficiency**

To fabricate the OBUS with a 15° conical angle, a steel rod mold with an axicon tip was machined. After coating the metal cone with candle soot (CS), the CS layer was transferred into PDMS and cured to form the OBUS device (**Supplementary Figure S2**). However, initial designs with a sharp conical tip led to heat accumulation and thermal damage of OBUS under high laser energy. To address this issue, the mold tip was polished to a slightly rounded shape. All references to OBUS refer to this optimized rounded-tip design.

The OBUS prepared has a diameter of 2.33 mm (**Fig. 1D inset**) and is ultralight, weighing only 2.1 mg. With the addition of the 3D-printed adapter, the total weight is 167.6 mg. This lightweight design highlights the potential of OBUS as a wearable device, providing its potential advantage over conventional ultrasound transducers for use in freely moving animals.

To characterize the optoacoustic properties, a laser with a wavelength of 1064 nm, pulse width of 2.2 ns, and fluence of 61 µJ/cm² was delivered to OBUS, generating a peak-to-peak pressure of 4.1 MPa at a distance of 2 mm from the OBUS surface (**Fig. 1D**). The frequency spectrum of the waveform, calculated using FFT, revealed a central frequency of 10.6 MHz and a -6 dB bandwidth of 250% (5–30 MHz). In **Fig. 1E and F**, spatial profiling of OBUS was conducted using a needle hydrophone, yielding lateral and axial resolutions of 152 µm and 1.93 mm at the focal point's full width at half maximum (FWHM), confirming the creation of an elongated acoustic field.

To optimize optoacoustic conversion efficiency, we measured the peak-to-peak pressure generated by OBUS using various PDMS pre-polymer to curing agent weight ratios. Previous studies have shown that the weight ratio of PDMS pre-polymer to curing agent affects the Young's modulus $E$ of pure PDMS(*40*). According to optoacoustic principles, the initial pressure $p_0$ generated in the material is directly proportional to the Gruneisen parameter $\Gamma$:

$$p_0 = \Gamma \cdot \mu_a \cdot F \qquad (2)$$

where $\mu_a$ is the absorption coefficient, $F$ is the optical fluence. The Gruneisen parameter $\Gamma$ is proportional to the bulk modulus $K$ in solid materials:

$$\Gamma = \frac{\beta}{\rho C_v \kappa} = \frac{\beta K}{\rho C_v} \qquad (3)$$

where $\beta$ is the thermal coefficient of volume expansion, $\rho$ is the mass density, $C_v$ is the principal heat capacity at constant volume, and $\kappa$ is the isothermal compressibility. The bulk modulus $K$ is further proportional to Young's modulus $E$:

$$K = \frac{E}{3(1-2\nu)} \qquad (4)$$

where $\nu$ representing the Poisson's ratio. Since the CS-embedded PDMS in OBUS is a composite material, we experimentally evaluated the impact of varying the weight ratio on optoacoustic conversion efficiency. OBUS was fabricated at weight ratios of 2:1, 4:1, 6:1, 8:1, and 10:1 to assess their impact on peak-to-peak pressure under identical laser energy conditions. As shown in **Fig. 1G**, the 8:1 ratio produced the highest acoustic amplitude, indicating optimal optoacoustic efficiency. This 8:1 ratio was used for the final characterization (**Fig. 1D-F**) and subsequent experiments.

**Transcranial efficiency and VTA maintenance of OBUS after skull aberration**

To assess the transcranial capability of OBUS, we simulated and compared its focus retention through a rat skull with that of a Gaussian beam. We performed simulations instead of experimental testing because, for a fair comparison, precisely targeting the same depth is straightforward in simulation but challenging to achieve in device fabrication. In the simulation, a real profile of a rat skull was scanned by ultrasound and imported, along with the OBUS profiles. A curved surface was designed to generate the Gaussian beam. By varying the diameters of the devices, both OBUS and Gaussian beams were designed to focus at a 4.8 mm depth, corresponding to the rat thalamus. This brain region was chosen for two reasons: Firstly, it aims to maximize differences in outcomes and demonstrate OBUS's ability to target deeper brain regions while preserving the desired shape. Secondly, the thalamus is a well-established target in brain studies due to its influence on cortical functions (*41*).

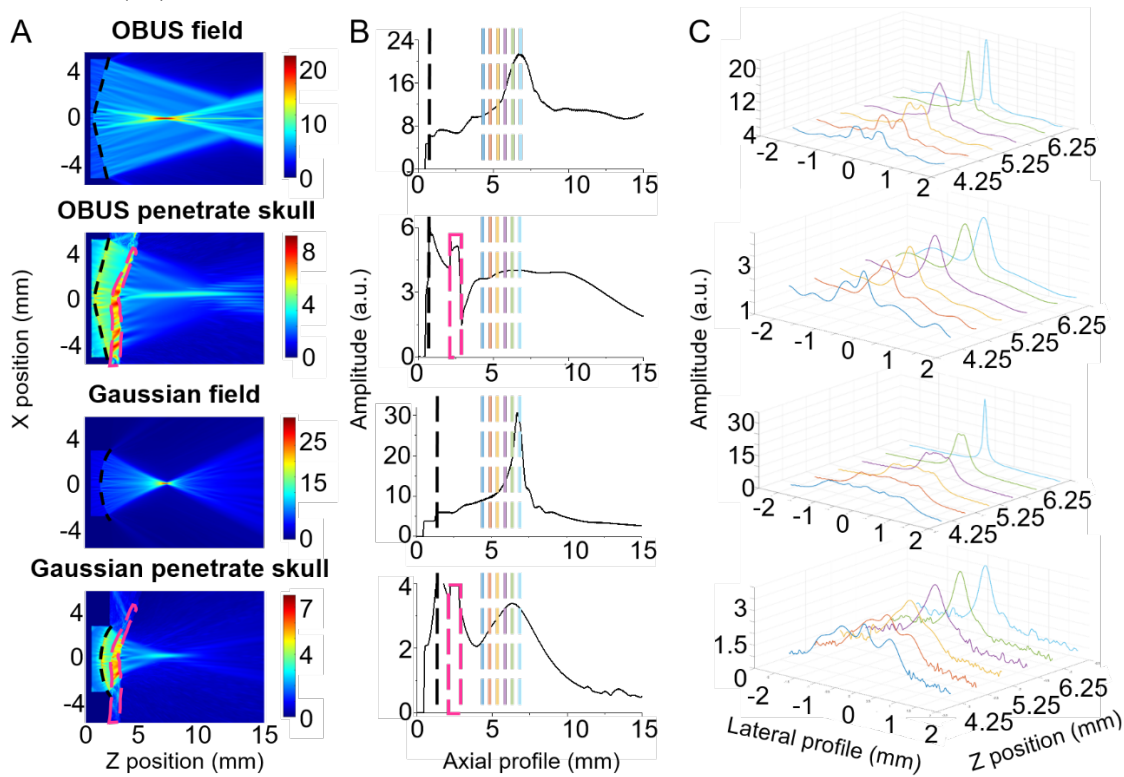

**Figure 2 Comparison of simulated acoustic fields between OBUS and Gaussian beam ultrasound.** (A) Simulated acoustic fields of OBUS and Gaussian beam ultrasound without and with interacting with the rat skull. Black dash lines: the surface of the emitters. Pink regions: the rat skulls. (B) Axial profile of the maximum pressure along the propagation direction. Black dash lines: the surface of the emitters. Pink rectangular: the position of the rat skull. The blue, red, yellow, purple, green, and light blue dash lines: the positions for the lateral profiles in (C). (C) Lateral profiles of the acoustic field at six positions: the intended focus and five preceding slices at 0.5 mm intervals.

We set both fields to a central frequency of 10 MHz with a 250% bandwidth, based on experimental data measured from OBUS. Water served as the propagation medium. Simulations of acoustic fields with and without the skull in place are shown in **Figure 2** for analysis. As shown in **Fig. 2A**, the presence of the skull induced aberrations in both OBUS and Gaussian beams. **Fig. 2B** shows that both focal points shifted 0.5 mm closer to the skull due to its high acoustic impedance. For reference, the simulation results of a standard acoustic Bessel beam are provided in **Supplementary Fig. S3**.

To further analyze the acoustic fields, we examined the lateral profiles at the intended focal point and at positions up to 2.5 mm anterior to it, with 0.5 mm increments, as shown in **Fig. 2C**. In the transcranial profiles, OBUS maintained an elongated focus after penetrating the skull, with a side lobe appearing immediately post-penetration. In contrast, the Gaussian beam exhibited a broad profile extending between 4.25 and 5.25 mm along the Z-axis, with an intensity remaining above -6 dB relative to the peak. This suggests that in transcranial neuromodulation with the Gaussian beam, tissue along the propagation path may receive exposure levels comparable to those at the intended target, potentially leading to unintended effects and excessive heat deposition. Conversely, OBUS demonstrated a well-defined VTA, maintaining the designed focal region while minimizing exposure to surrounding areas. These results underscore OBUS's capability for precise neuromodulation with reduced off-target effects in deep-brain transcranial applications.

To quantify VTA maintenance, we calculated the ratio between simulated peak intensity after and before skull penetration to assess transcranial efficiency. We also measured axial and lateral resolutions pre- and post-skull penetration, defined as the FWHM values, and calculated the percentage of change for comparison. **Table 2** summarizes these results: OBUS achieved a transcranial efficiency of 18.7%, surpassing the Gaussian beam by 70%. Additionally, OBUS showed only a 40.5% axial change, whereas the Gaussian beam experienced a drastic 566.7% increase. Similarly, the percentage change in lateral resolution for OBUS was 74% of that observed in the Gaussian beam. These findings underscore OBUS's superior ability to preserve VTA, enabling precise neuromodulation while minimizing unintended exposure. This underscores its potential for targeted transcranial applications. The corresponding data for the standard acoustic Bessel beam are provided in **Supplementary Table S1**.

| Performance parameters | OBUS | Gaussian Beam |
|---|---|---|
| Peak intensity transcranial efficiency | 18.7 % | 11.0% |
| Axial resolution (mm) – pre-skull penetration | 6.48 | 1.05 |
| Axial resolution (mm) – post-skull penetration | 2.63 | 5.95 |
| Percentage of change in axial resolution | 40.5% | 566.7% |
| Lateral resolution (mm) – pre-skull penetration | 0.20 | 0.10 |
| Lateral resolution (mm) – post-skull penetration | 0.93 | 0.63 |
| Percentage of change in lateral resolution | 462.5% | 625.0% |

**Table 2. Summary of transcranial efficiency on peak intensity and changes in axial and lateral resolution before and after skull penetration for OBUS and Gaussian beam ultrasound.**

**Elongated stimulation volume delivery in vivo with OBUS**

To evaluate OBUS's ability to induce cellular responses with a targeted VTA in vivo, we used c-Fos immunofluorescence staining. c-Fos has been widely used as a marker of neural activation following repeated stimulation. We conducted in vivo stimulation in adult C57BL/6J mouse (N =1), exposing the skull and applying ultrasound gel for optimal acoustic coupling **(Fig. 3A)**. Using stereotaxic coordinates (AP: -0.5, ML: 1.5), we aligned the ultrasound focus with the motor cortex, maintaining a 0.5 mm gel gap between OBUS and the mouse head to prevent heat accumulation on the tissue. For stimulation, OBUS was operated to produce a peak-to-peak pressure of 4.6 MPa at the focus, delivering 500-pulse trains every 2 seconds for a duration of 30 minutes **(Fig. 3B)**. Following stimulation, the mice rested for 1 hour to facilitate peak c-Fos expression. Subsequently, the brains were extracted, fixed, and sectioned horizontally for VTA measurement and immunofluorescence analysis. To aid stimulation site identification, extracted brains were punctured at Bregma and the contralateral untreated side (**Supplementary Fig. S4**).

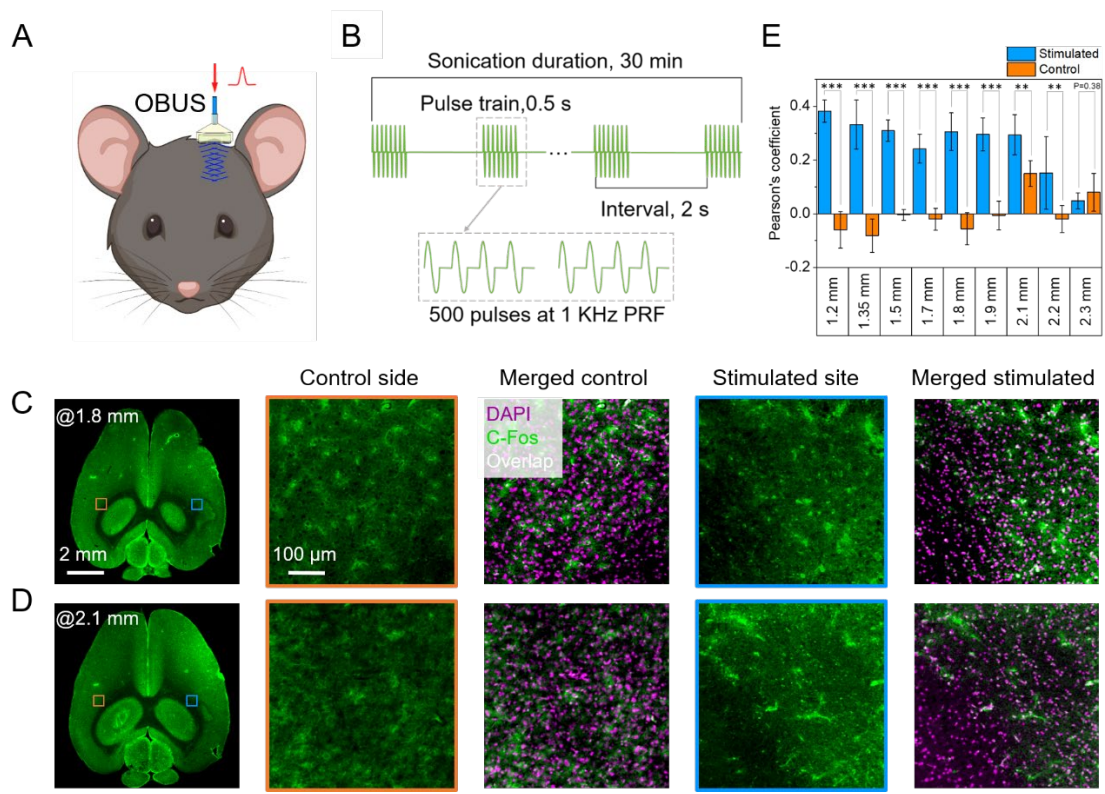

**Figure 3 Elongated stimulation volume delivery in vivo with OBUS.** (A) Schematic of the experimental setup for OBUS stimulation. (B) Schematic representation of the pulse train delivered to induce c-Fos protein expression. (C) and (D) Representative images of the brain slice at 1.8 mm and 2.1 mm, respectively. (E) Comparison of Pearson's coefficient between stimulated site and control sites at different depths. Position 0 mm is at the surface of OBUS.

To confirm cellular activation, we performed DAPI and c-Fos staining and conducted confocal imaging of both markers at various depths (**Supplementary Figure S5-6**). Brain slices were analyzed with the brain surface as the reference point to determine the depth of the VTA. Brain slices shallower than 1.2 mm were excluded from analysis, as they either fell outside the OBUS field or contained the stimulation site near the edge, making data collection impractical. Some tissue damage was observed at the edges of the brain slices due to cuts during extraction. Representative images at 1.8 mm (**Fig. 3C**) and 2.1 mm (**Fig. 3D**) showed concentrated signals at the targeted region. The stimulated site exhibited strong c-Fos signals, while the contralateral control side showed minimal signal. Overlay images confirmed the colocalization of c-Fos (green) and DAPI (magenta) as white signals, indicating cellular activation specifically induced by OBUS stimulation.

For quantitative depth analysis of the VTA, ImageJ software was used to perform unbiased colocalization analysis for c-Fos and DAPI signals through automated image thresholding and calculation of the Pearson's correlation coefficient, which ranges from -1 (negative correlation) to 1 (positive correlation). Slight misalignments in zoomed images and asymmetry between the stimulated and control groups likely resulted from manual adjustments of the confocal field. To address these issues, five smaller regions of interest (ROIs) were selected within the targeted region in both stimulated and control sites. The following criteria were applied to standardize selection: (1) Avoidance of vessels, which were identified by their larger size relative to nuclei to prevent overestimation of colocalization; (2) Ensuring overlap with the targeted site (white dashed circle in **Supplementary Figure S5-6**); and (3) Maintaining symmetry between ROIs at stimulated and control sites, as indicated by yellow boxes in **Supplementary Figure S5-6**. An unpaired t-test was performed to compare Pearson's coefficients at various depths, with results shown in **Fig. 3E**. Notably, significant differences in Pearson's coefficient between the stimulated and control groups were observed up to a depth of 2.2 mm.

When an additional 0.6 mm was added to accounted for the gap of ultrasound gel to the 2.2 mm depth, the decay pattern of the c-Fos signal closely matched the axial acoustic profile, with the Pearson's coefficient for the c-Fos signal peaking at 2.2 mm and gradually decaying to 2.8 mm (**Supplementary Figure S7**). The axial acoustic profile measured by the hydrophone (**Fig. 1F**) peaked around 2 mm and decayed to 40% at 3 mm. This alignment confirmed that the depth of the c-Fos signal corresponded to the OBUS VTA.

**Electrophysiological responses induced in vivo by OBUS**

After confirming OBUS's ability to target the VTA through c-Fos staining, we evaluated its potential to elicit localized electrophysiological responses in vivo in C57BL/6J mice. Craniotomy was performed on anesthetized mice, and OBUS was aligned with the somatosensory cortex at stereotaxic coordinates (AP: -1.5, ML: 2). A 3D-printed pointer indicated the VTA lateral center, ensuring precise alignment with the recording electrode (**Supplementary Figure S8**). A 16-channel electrode (NeuroNexus) was then accurately inserted into the brain using a 3D stereotactic frame (**Fig. 4A**). To minimize heat accumulation, OBUS was placed approximately 0.5 mm away from the brain surface.

For neuromodulation experiments in mice, OBUS was applied at peak-to-peak pressures of 2.8 MPa, 3.7 MPa, and 4.1 MPa with a 1 kHz pulse repetition frequency. Each stimulation consisted of a 10-second pulse train followed by a 20-second rest period within a 1-minute cycle. Electrophysiological recordings lasted 3 minutes, comprising 1-minute baseline before stimulation, 1-minute during OBUS stimulation, and 1 minute after stimulation (**Fig. 4B**). Each pressure level was tested in triplicate in three mice (N = 3). Raw signals were recorded using cortical electrodes, and local field potentials (LFPs; 0.5–100 Hz) were extracted for analysis.

A representative LFP trace and its corresponding spectrogram during 4.1 MPa OBUS stimulation are shown in **Fig. 4C**, demonstrating increased LFP frequency and amplitude both during and after stimulation. Data from additional trials are provided in **Supplementary Fig. S9-11**. **Fig. 4D** shows the power spectral density (PSD) calculated at each pressure level in one representative mouse. Notably, PSD analysis revealed increased power in the 10–50 Hz range at all three pressure levels, corresponding to beta and gamma bands. Although 2.8 MPa produced the largest PSD increase, high variability was observed, likely due to the limited LFP signal recorded (**Supplementary Fig. S9**). In contrast, PSD increases at 3.7 and 4.1 MPa were more consistent and reproducible, with smaller error bars, indicating improved reliability with increasing pressure.

**Fig. 4E** shows the normalized PSD (During normalized to Before and After normalized to Before) across all mice (N = 3). A paired t-test on raw data (before normalization) at 46.9 Hz—the frequency with the largest PSD increase at 4.1 MPa—revealed no significant difference at 2.8 MPa but significant increases at 3.7 and 4.1 MPa. Post-stimulation PSD ("After") also showed significant elevation compared to baseline at all pressure levels.

These results indicate that OBUS effectively modulates neural activity. Higher pressure levels produced more significant increases in brain activity during stimulation; however, the sustained post-stimulation effects were consistent across all pressures, suggesting that the persistence of neuromodulatory effects is pressure-independent.

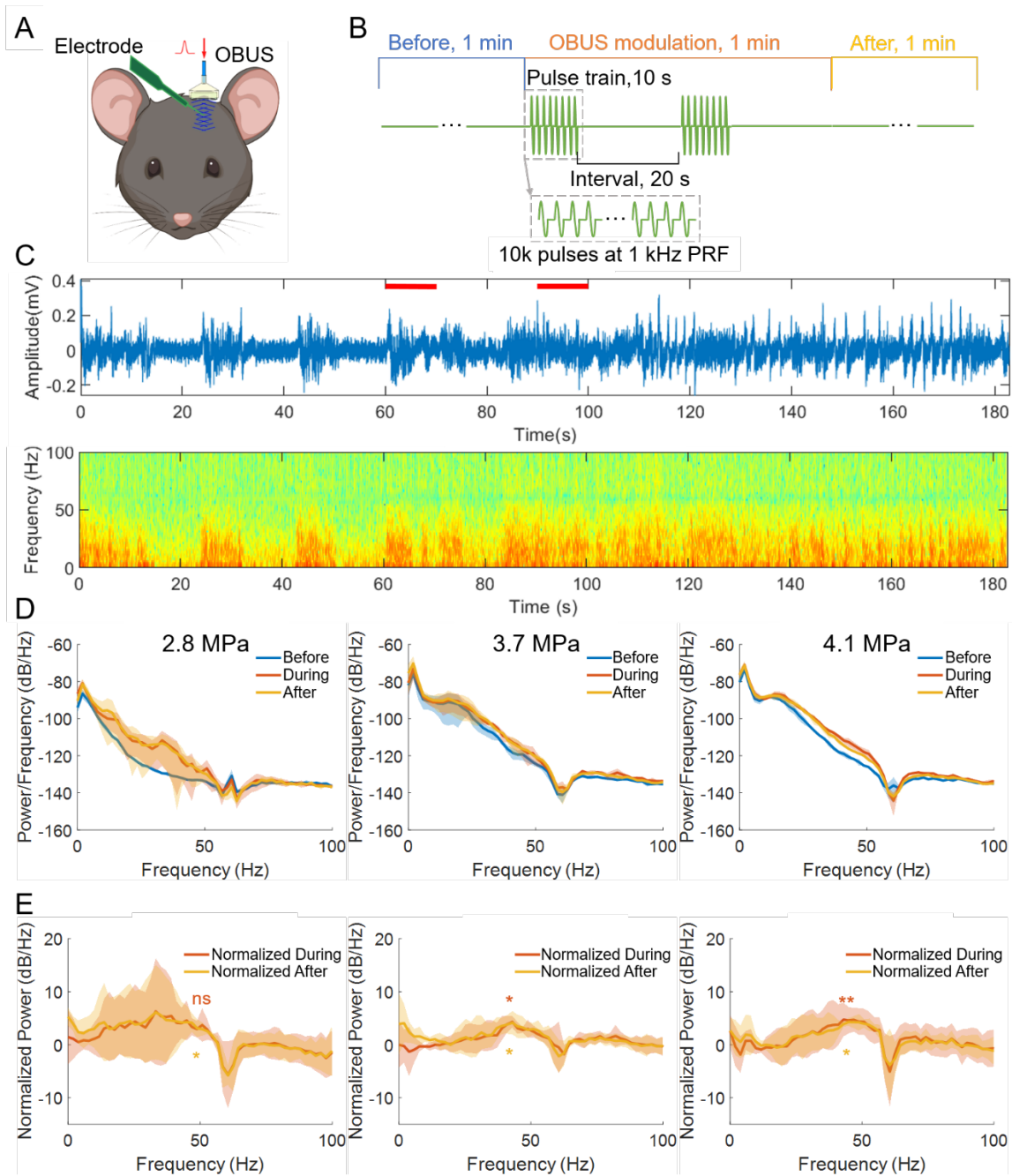

**Figure 4 LFP responses induced by OBUS modulation.** (A) Schematic of the in vivo OBUS stimulation and recording setup. (B) Schematic representation of the pulse train delivered to stimulate LFP responses. (C) Representative LFP signals and corresponding frequency spectrogram during OBUS stimulation at 4.1 MPa, initiated at t = 60 s and 90 s with a 10 s duration. Red bars indicate laser burst timing. (D) Power spectral density (PSD) plots for 2.8MPa, 3.7 MPa, and 4.1 MPa (left to right), averaged over three repeats (N = 1 mouse). (E) Normalized power spectral density (PSD) plots for 2.8 MPa, 3.7 MPa, and 4.1 MPa (left to right). The "during" and "after" PSD values were normalized to "Before" and then averaged across three animals (N = 3 mice). Paired t-tests were conducted to compare "Before vs. During" and "Before vs. After".

To verify that the observed LFP signals were specifically induced by OBUS, we conducted two control experiments. First, we set the laser output to 0 mW to eliminate potential electrical interference. No increase in LFP amplitude or frequency was observed in the signal trace (**Supplementary Fig. S12A**), and no change in the PSD was detected (**Supplementary Fig. S12B**). These results confirm that the neural responses observed in prior experiments were not artifacts of electrical interference. Second, we repositioned the OBUS transducer to the contralateral side of the S1 cortex and applied stimulation at 4.2 MPa to evaluate whether the responses were mediated by cochlear activation. No significant changes in LFP amplitude or frequency were detected (**Supplementary Fig. S13A**), and PSD remained unchanged (**Supplementary Fig. S13B**). These findings indicate that OBUS-induced modulation is spatially localized and results from direct interaction with brain tissue, rather than indirect activation via the auditory pathway.

**Blood oxygenation level dependent (BOLD) responses elicited by OBUS modulation**

To non-invasively confirm the induction of brain activity by OBUS, we monitored the percent change in BOLD signal in the somatosensory cortex of rats (N=2) using fMRI. As shown in **Fig. 5A**, we conducted a positive control by applying electrical stimulation to the right hind paw for 15 seconds every minute for seven stimulation blocks, with the OBUS device positioned but inactive. For OBUS stimulation, the device was activated with a peak-to-peak pressure of 4.9 MPa at a 1 kHz pulse repetition frequency for 15 seconds every minute, across seven modulation blocks. The representative time course of the BOLD signals for both conditions are shown in **Fig. 5B**, which indicates a clear correlation between the stimulus activation and the increase in signal response over time. The BOLD signal maps were obtained as a percent change from the unstimulated baseline and averaged over the stimulation blocks to provide an average signal over the 1-minute on/off modulation block. The functional activation maps (**Fig. 5C**) reveal spatially distinct activation patterns: electrical stimulation of the right hind paw evoked responses in the left somatosensory cortex, while OBUS stimulation consistently elicited activation directly beneath the device. Both robust positive and negative modulation was observed during OBUS modulation. The averaged data over the stimulation blocks for each rat (**Fig. 5D**) show that BOLD signals followed both stimulation paradigms, peaking at 15 seconds and returning to baseline post-stimulation. The peak BOLD responses from OBUS were comparable to those from electrical stimulation, indicating that OBUS may serve as an effective alternative neuromodulation method.

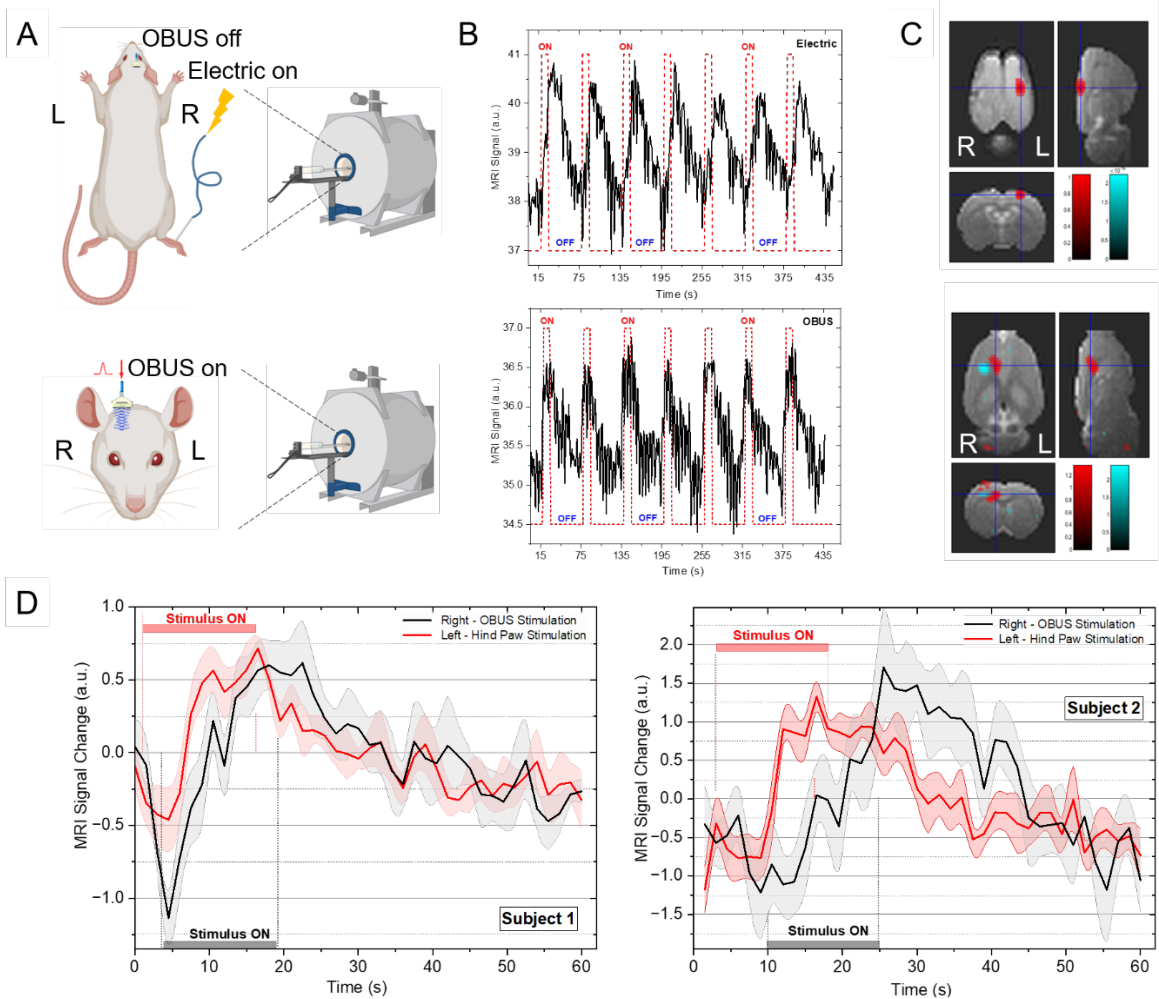

**Figure 5 fMRI BOLD signal in response to OBUS modulation in the somatosensory cortex.** (A) Experimental design showing electrical stimulation applied to the right hind paw (positive control) and OBUS modulation with the device positioned over the cortex. (B) Representative BOLD signal time courses for both conditions, aligned with the stimulation period. Top: electrical stimulation: Bottom: OBUS modulation. Red dashed lines indicate stimulation onset and duration. (C) Representative functional activation maps averaged across stimulation blocks, illustrating contralateral cortical activation from electrical hind paw stimulation (top) and localized modulation beneath the OBUS (bottom). (D) Averaged BOLD responses for individual subjects over stimulation blocks showing consistent temporal dynamics, with signal peaks at 15 seconds followed by a return to baseline after stimulation ends.

### Safety evaluation of in vivo OBUS neural modulation

To ensure the safety of in vivo OBUS stimulation in mice, we evaluated both the mechanical index (MI) and temperature rise associated with the stimulation parameters.

The MI quantifies the potential for tissue damage due to rarefactional pressure. For our experiments, we set a peak-to-peak pressure limit of 5 MPa for subsequent animal studies. Based on the negative-to-positive peak pressure ratio shown in **Fig. 1D**, we estimated the negative peak amplitude to be 3.05 MPa, resulting in an MI of 0.93—well below the FDA threshold of 1.9 (*42*), indicating safe operation at 5 MPa.

To assess thermal deposition, we measured temperature rise in water with a thermocouple, as direct temperature measurements in brain tissue are impractical. Measurements were taken at distances up to 2 mm from the OBUS surface under two conditions: continuous wave (CW) mode, which produced the maximum temperature rise among all patterns, and a 25% duty cycle, the condition used for c-Fos immunofluorescence studies. Each recording lasted 20 seconds to ensure thermal equilibrium was reached. **Supplementary Fig. 14A** shows the temperature profile at 0.5 mm, approximating the thermal exposure at the tissue surface. **Supplementary Fig. 14B** shows temperature variations across different depths. At the OBUS surface, temperature increases ranged from 1.2 K to 2.2 K, but at 0.5 mm, the rise dropped to approximately 0.5 K. Given that temperature increases remained below 1 K, OBUS is considered safe for in vivo use, and the minimal thermal rise confirms that recorded stimulation effects are unlikely to be thermally induced. In summary, OBUS is safe for in vivo neural modulation, with a low MI and negligible temperature rise during stimulation.

**Discussion**

In this study, we designed and developed an OBUS device for non-invasive targeting of column-shaped brain regions. We characterized the spatial profile of the OBUS-generated ultrasound field and demonstrated its enhanced transcranial penetration capability compared to a conventional Gaussian beam based photoacoustic emitter. To our knowledge, this is the first report achieving non-invasive neuromodulation with precise control over the shape of the VTA. Notably, OBUS outperformed conventional Gaussian beams in transcranial applications, as confirmed through simulations using rat skulls. This makes OBUS a more effective neuromodulation technique compared to conventional ultrasound neuromodulation. To validate OBUS's targeting capability, we performed immunofluorescence imaging for c-Fos protein, revealing a VTA depth of 2.2 mm in mouse brains. Additionally, we recorded successful electrophysiological and BOLD responses to OBUS stimulation in mice and rats. We confirmed the device's safety through mechanical index (MI) and temperature rise measurements with the stimulation pattern designed for in vivo experiments. This comprehensive evaluation demonstrates OBUS's potential as a non-invasive neuromodulation tool with advanced spatial control and safety features.

It should be noted that the c-Fos data in **Figure 3** were acquired using a cranial window rather than a fully transcranial method. We have obtained c-Fos positive data with an intact skull with another mouse, however, repeated experiments with the intact skull had a lower success rate. In the craniotomy experiments, the VTA identified by the c-Fos–positive region exhibited a lateral profile exceeding 1 mm (**Supplementary Figure 5**, stimulated side at 1.2 mm depth), suggesting that stimulation was driven by the entire OBUS beam rather than only the focal area. Given that the lateral width of the OBUS focal area is approximately 150 μm wide—encompassing only a few neurons—it was difficult to spatially localize activated regions in whole-brain slices. To overcome this limitation, we increased the acoustic pressure to engage the full OBUS beam, allowing us to more reliably identify the stimulated area. As a result, while c-Fos expression serves as a reliable indicator of stimulation depth, it has limited utility for resolving lateral activation patterns.

Optoacoustic technology offers distinct advantages over conventional ultrasound techniques for generating specific acoustic fields, such as Bessel beams. Bessel beams are commonly produced using three methods in traditional ultrasound systems: acoustic lenses (*43*, *44*) or spiral phase plates (*45*) on a planar ultrasound transducer, acoustic transducer arrays (*31*), and specialized transducers such as ring arrays (*46*). However, these methods face inherent limitations. Acoustic lenses, for instance, suffer from energy losses due to reflections at two interfaces, leading to reduced acoustic intensity. This issue is exacerbated when materials with high acoustic impedance mismatches are used. For example, with aluminum, only about 3% of the acoustic energy is transmitted (*43, 47*). Even lower impedance materials like epoxy still reflect approximately 20% of incident energy (*44*). Furthermore, while ultrasound arrays and specialized transducers enable more precise control of the wavefront, they are often complex and require advanced equipment and calibration, limiting their practicality. Optoacoustic modality overcomes many of these challenges by employing a direct emitting surface to generate acoustic fields. This approach bypasses the need for intermediate components, such as lenses, reducing energy loss and improving acoustic transmission efficiency. Furthermore, compared to the complexity of transducer arrays and ring transducers, a single-element emitter in optoacoustic systems offers a simpler, more accessible, and cost-effective method for generating the desired VTA for brain research. This simplicity, combined with its efficiency, makes the optoacoustic modality a practical and innovative solution for creating tailored acoustic fields for neuromodulation.

We selected the Bessel beam for shaping the VTA due to its advantageous properties for non-invasive brain stimulation. First, its greater penetration depth, compared to Gaussian beams from conventional curved acoustic surfaces, is ideal for targeting column structures in the brain. Second, the Bessel beam's self-healing property enables it to recover its shape after encountering small obstacles, such as blood vessels, within the brain's heterogeneous structure (*31*). This resilience, combined with OBUS's superior shape-maintaining capabilities, ensures reliable performance in complex brain environments. Third, simulations showed that Bessel beams maintain focal intensity despite emitter size variations, supporting miniaturization for compact optoacoustic devices. The acoustic intensity of other optoacoustic devices, such as SOAP, scales with its size due to the material's fixed laser energy tolerance. In contrast, OBUS maintains consistent intensity despite variations in its depth of focus. Lastly, while standard Bessel beams have lower localized intensity than Gaussian beams (*48*), our polished round tip design combines the strengths of both, increasing peak intensity within the VTA by 51.6% (**Supplementary Figure 3**) while extending the beam profile and reducing heat accumulation during skull penetration. These attributes make the Bessel beam, particularly in its modified form, a powerful and practical tool for non-invasive brain modulation.

In comparing OBUS to optogenetics, the latter represents a well-established technology with its unparalleled selectivity for stimulating specific brain subregions (*18*) but presents practical challenges for four reasons. First, optogenetics requires significant expertise, as viral vector optimization for specific brain regions can be complex and time-intensive. Multi-site, low-pressure viral injections demand advanced skills and careful training (*19*). Additionally, maintaining cranial windows or implanted lenses on animal skulls requires meticulous care to prevent infections. In contrast, OBUS, an optoacoustic neuromodulation device with a portable laser, is easy to distribute and implement across labs, as shown by the successful use of optoacoustic stimulation devices shared with collaborators. Optoacoustic neuromodulation also avoids the need for invasive surgeries and associated cranial maintenance, making it an accessible tool for fundamental research. Secondly, while optogenetics can require up to 3 months for viral transfection (*19*), leading to significant delays in animal studies, optoacoustic neuromodulation allows research to start immediately once animals are available, greatly accelerating research timelines and improving workflow efficiency. Thirdly, while blue light in optogenetics retains only 1% of its energy at a depth of 0.9 mm(*19*), OBUS's ultrasound-based stimulation reaches a depth of 2 mm, covering brain sub regions like the ODC across the cortex's full thickness (~2 mm). Lastly, optogenetic stimulation often involves a ~50% duty cycle, raising thermal accumulation risks. In contrast, optoacoustic neuromodulation applies single ultrasound cycles within microseconds at kHz frequencies, yielding a much lower duty cycle and significantly reducing thermal risks. Overall, OBUS provides potential as a miniaturized and light-weighted platform for neural stimulation in freely-moving animals with high penetration depth and controllable VTA.

## Materials and Methods

### Fabrication of OBUS device

To fabricate the desired conical mold, a steel rod (McMaster-Carr, 8890K1) was machined and its tip polished into a curved shape to prevent heat accumulation and potential thermal damage to the OBUS tip. The mold was then exposed to a paraffin wax candle flame for 10–15 seconds to achieve a uniform carbon soot (CS) coating. Next, the coated mold was immersed in a degassed PDMS mixture with base-to-curing agent ratios of 2:1, 4:1, 6:1, 8:1, and 10:1 to optimize optoacoustic conversion efficiency. The immersion took place within a 2.33 mm diameter metal mold to ensure precise shaping. Finally, the coated mold was cured at 110°C for 15 minutes to solidify the structure and complete sample formation.

### Characterization of the ultrasound field generated by OBUS

A diode-pumped laser (Onda 1064 nm, Bright Solution) emitting 2.2 ns pulses at a 1064 nm wavelength was used to generate optoacoustic signals with the OBUS. The laser beam was modulated at a 1 kHz repetition rate via a function generator (33220A, Agilent) and a pulse generator (9214-TZ50, Quantum Composers). To minimize laser leakage, an optical chopper was integrated to ensure consistent 1 kHz pulse delivery. Laser coupling was achieved using a 400 μm multimode fiber (FT400EMT, Thorlabs), and the numerical aperture (NA) of the output beam was characterized using a CMOS camera (MU1000, Amscope). A custom 3D-printed adapter was designed to facilitate uniform illumination of the OBUS device.

For ultrasound pressure and waveform characterization, a measurement system comprising a needle hydrophone (NH0040, optimized for 5–40 MHz, Precision Acoustics), a submersible preamplifier, and a DC coupler was employed. The signal was further amplified using an ultrasonic pulse-receiver (Model 5073PR, Olympus) and recorded with a digital oscilloscope (DS4024, Rigol) after four-time averaging. The 40 μm needle hydrophone was utilized to acquire waveform and pressure data for OBUS samples fabricated with varying PDMS base-to-curing agent ratios. Additionally, the hydrophone was mounted on a 3D manipulator to obtain spatial profiles of the OBUS-generated acoustic field.

### Simulation of the ultrasound field generated by OBUS

The ultrasound field generated by OBUS was simulated using the open-source k-Wave toolbox in MATLAB R2021a (MathWorks, MA). The simulation was performed in 2D, leveraging rotational symmetry, and focused solely on acoustic wave propagation, disregarding light absorption and photoacoustic conversion. Material-specific density and acoustic velocity values were assigned accordingly. In the 2D setup, a PDMS block with various surface profiles was positioned at the water-air interface, with water as the propagation medium and air as the backing material, to compute the resulting acoustic field.

To optimize the conical angle prior to OBUS fabrication, the acoustic wave's central frequency was set to 15 MHz with a 200% bandwidth according to the previous study (*38*). To accurately model the acoustic field produced by the tip-polished OBUS, the emitter's surface profile was extracted from a photograph and imported into the simulation. Based on experimental OBUS characterization, the acoustic wave's central frequency was set to 10 MHz with a 200% bandwidth. To study skull-induced aberrations, a rat skull profile obtained via photoacoustic tomography was incorporated into the simulation.

**In vivo stimulation on mice via OBUS**

All experimental procedures adhered to the relevant guidelines and ethical regulations for animal research, as approved by the Institutional Animal Care and Use Committee of Boston University (PROTO201800534). Adult C57BL/6J mice (14–16 weeks old) were anesthetized with 5% isoflurane in an oxygen chamber before being secured in a standard stereotaxic frame. Anesthesia was maintained at 1.5–2% isoflurane, with depth monitored via the tail pinch reflex. A heating pad was placed beneath the animal to maintain body temperature. The targeted brain region was prepared by removing the fur, and the OBUS device, mounted on a 3D-printed holder, was aligned with the mouse somatosensory cortex (AP: -1.5 mm, ML: 2 mm). To minimize thermal deposition, the OBUS was positioned approximately 0.5 mm above the tissue. The acoustic coupling was maintained by filling with ultrasound gel. The laser pulses at a 1 kHz repetition rate were delivered to the OBUS via an optical fiber.

To induce c-Fos protein expression, OBUS stimulation was applied in a pulse train with a 25% duty cycle (0.5 s laser on, 1.5 s laser off) for 30 minutes. To evoke local field potentials (LFPs), 10-second OBUS stimulations consisting of 10,000 optoacoustic wave cycles were delivered every 20 seconds.

**Immunofluorescence staining and imaging of mouse brain**

After the stimulation session, mice were allowed to rest for one hour to maximize c-Fos expression. Euthanasia and transcardial perfusion were then performed using 1× phosphate-buffered saline (PBS, pH 7.4; Thermo Fisher Scientific), followed by fixation with 10% formalin. To facilitate stimulation site identification, extracted brains were carefully punctured at Bregma and the contralateral untreated site using a 25G needle. The brains were post-fixed in 10% formalin for 24 hours before being transferred to 1× PBS. Horizontal sections (100 μm thickness) were obtained using an oscillating tissue slicer (OST-4500, Electron Microscopy Sciences) to assess stimulation depth. Brain slices were then transferred back into 10% formalin for an additional 24-hour fixation.

For immunostaining, slices were blocked in 5% bovine serum albumin (Sigma-Aldrich) in PBS for 30 minutes at room temperature and permeabilized with 0.2% Triton X-100 (Bio-Rad) in PBS for 10 minutes. They were incubated overnight at 4°C with an anti-c-Fos rabbit antibody (2 μg/mL; Cell Signaling Technology), followed by incubation in the dark at 4°C for 2 hours with Alexa Fluor 488 goat anti-rabbit IgG (1 μg/mL; Thermo Fisher

Scientific) and DAPI (4 μg/mL; Thermo Fisher Scientific) for nuclear staining. Between each step, slices were rinsed four times for 5 minutes each in 0.2% Tween 20 (Tokyo Chemical Industry) in PBS.

Fluorescent images were acquired using an FV3000 Confocal Laser Scanning Microscope (Olympus) with separate excitations at 405 nm for DAPI and 488 nm for c-Fos to minimize spectral cross-talk.

**Electrophysiological recording and signal processing**

A multichannel neural probe (A1x16-Poly2-8mm-100s-177-A16, NeuroNexus) with ~1 MΩ impedance was surgically implanted to mice brains to record electrophysiological signals at the stimulation site. Neural signals were acquired at a 20 kHz sampling rate using a 16-channel headstage (Intan Technologies, Part #C3334) and digitized by a 512-channel controller (Intan Technologies, Part #C3004).

To synchronize laser activation with neural recordings, a trigger signal from the function generator was connected to an analog port on the recording system. For probe implantation, the anesthetized mouse underwent craniotomy. Alignment between the probe tip and OBUS center was achieved using a 3D micromanipulator. The OBUS was positioned at the stimulation site with ultrasound gel coupling, and laser delivery was precisely controlled using a mechanical shutter (Thorlabs Inc., SH1).

Data processing was performed in MATLAB using custom scripts. Raw extracellular recordings were bandpass-filtered between 0.5–100 Hz to extract local field potential (LFP) signals. Spectrograms of LFP activity were generated using a 512-point fast Fourier transform (FFT) with 50% overlap, providing a frequency resolution of 1.95 Hz. Power spectral density (PSD) before, during, and after OBUS stimulation was computed using Welch's method with a 512-point window and 50% overlap.

**Functional magnetic resonance imaging (fMRI) acquisition and processing**

All functional MRI experiments and animal handling procedures were approved by the Brigham and Women's Hospital Institutional Animal Care and Use Committee and conducted in compliance with the Office of Laboratory Animal Welfare and the Association for Assessment and Accreditation of Laboratory Care regulations. Healthy male Sprague–Dawley (SD) rats (~400 g, Charles River Laboratories, n = 2) were used. Body temperature and breathing rate were continuously monitored throughout the experiments.

To enhance acoustic transmission, a 5-mm diameter cranial window was surgically created by removing the skullcap (AP: -1.9 mm, ML: 2.2 mm) while preserving the dura matter. The exposed brain was hydrated with 0.9% NaCl saline and covered with a cotton swab. Rats were anesthetized via intraperitoneal injection of ketamine (80 mg/kg) and xylazine (10 mg/kg), and petroleum jelly (Vaseline, Unilever, NJ, USA) was applied to the eyes to prevent dehydration.

MRI scans were performed using a horizontal bore 7.0-T Bruker BioSpec® USR scanner (Bruker Corporation, Billerica, MA) with a 20 mm inner diameter multipurpose receive-only surface coil over the rat's head. Rats were placed in the prone position with their heads immobilized in a nose cone for isoflurane and oxygen delivery. The OBUS was positioned at the center of the cranial window with ultrasound gel coupling. Anesthesia was maintained with a combination of medetomidine (Dexdomitor, 0.025 mg/kg) and light isoflurane flow (*49*).

The functional imaging paradigm consisted of alternating 15-second "ON" stimuli and 45-second "OFF" periods over 7 minutes and 30 seconds. Ultrasound-evoked neuronal activity was validated by observing bilateral activation of the hind limb region in the somatosensory cortex following electrical stimulation of the hind paws. Four experimental conditions were tested in the same animal: trial 1: bilateral electrical stimulation of the hind paws only without OBUS device in place (15s ON, 45s OFF); trial 2: OBUS stimulation only with OBUS device in place (15s ON, 45s OFF); trial 3: bilateral electrical stimulation of the hind paws with OBUS in place (15s ON, 45s OFF); and trial 4: bilateral electrical stimulation of the hind paws and OBUS stimulation with OBUS in place (15s ON, 45s OFF). Hind paw stimulation consisted of sub-threshold electrical pulses (5 Hz) delivered during the "ON" period to activate the primary somatosensory cortex (S1) over six stimulation blocks.

fMRI data were acquired using a 2D single-shot gradient-echo EPI sequence ($0.5 \times 0.5 \times 1.0$ mm resolution, $64 \times 64$ matrix, 18 axial slices, 0.2 mm slice gap, TR = 1500 ms, TE = 18 ms). Prior to the functional image, a main field homogeneity was optimized using the Bruker MAPSHIM protocol, and an anatomical image was acquired with a T2-weighted RARE sequence ($0.469 \times 0.469 \times 0.5$ mm resolution, $64 \times 64$ matrix, 50 axial slices, no slice gap, TR = 5367.7 ms, TE = 37.5 ms) was used for anatomical imaging. For hind paw stimulation, pairs of 30-gauge needles were inserted into the second and fourth-digit pads of each hind paw. Electrical pulses (300 ms, 6 Hz) were delivered via a TENS unit (TU 7000, Tensunits.com, Largo, Florida), with voltages adjusted to 700 mV (~2 mA current) per hind paw, confirmed via oscilloscope measurements.

Data preprocessing was performed using SPM (2014) software and custom MATLAB scripts. T2-weighted anatomical images were segmented in SPM12 using the template of Valdés-Hernández et al. (*50*). The EPI images were first preprocessed, including slice time corrected, realigned, co-registered to the anatomical image, normalized to the template space, and spatially smoothed with a Gaussian filter of 0.8 x 0.8 x 0.8 mm FWHM. BOLD signals were analyzed as percent change from non-stimulated baseline, averaged across stimulation blocks to provide an average signal over the 1-min on/off stimulation block. To mitigate motion-related artifacts, a temporal band-pass filter (0.006 – 0.25 Hz) was applied, and motion traces and the average time signal from a white matter mask were regressed out. Activation was quantified using two metrics: Mean BOLD Signal Change: Computed over a defined ellipsoidal cortical region of interest (ROI: $1.75 \times 2.5$ mm in the cortical plane, $\times 2.25$ mm through the cortex). Voxel-Based Activation Count: The number of voxels exceeding a 0.5% BOLD signal change within an expanded S1 ROI ($3.5 \times 4.5 \times 3.25$ mm). Data were averaged within subjects and presented as mean ± standard error.

**Temperature measurement of OBUS under stimulation pattern**

A thermocouple (DI-245, DataQ) was used to measure the temperature profile at both the acoustic focus and OBUS surface in water. It was incrementally repositioned along the focus region in 0.5 mm steps to capture temperature variations. Each recording lasted 20 seconds to ensure thermal equilibrium, with a sampling rate of 2000 Hz. Temperature measurements were conducted with the OBUS delivering an acoustic pressure of 5 MPa, consistent with in vivo conditions. The OBUS was operated in continuous wave (CW) mode and at 25% duty cycles.

**Statistical Analysis**

Acoustic waveforms and electrophysiological traces were plotted using Origin 2020. Frequency spectrum analysis was performed using fast Fourier transform (FFT) in MATLAB R2021a. Data are presented as mean ± standard error of the mean (SEM). Immunofluorescence images were analyzed using ImageJ. Additional data analysis details are provided in the respective method sections.

**References**
1. B. A. Vogt, Pain and emotion interactions in subregions of the cingulate gyrus. *Nat Rev Neurosci* **6**, 533–544 (2005).

2. K. C. Berridge, Affective valence in the brain: modules or modes? *Nat Rev Neurosci* **20**, 225–234 (2019).

3. R. Martínez-Fernández, L. Zrinzo, I. Aviles-Olmos, M. Hariz, I. Martinez-Torres, E. Joyce, M. Jahanshahi, P. Limousin, T. Foltynie, Deep brain stimulation for Gilles de la Tourette syndrome: A case series targeting subregions of the globus pallidus internus. *Movement Disorders* **26**, 1922–1930 (2011).

4. J. E. Hoover, P. L. Strick, Multiple output channels in the basal ganglia. *Science* **259**, 819–821 (1993).

5. D. L. Adams, L. C. Sincich, J. C. Horton, Complete Pattern of Ocular Dominance Columns in Human Primary Visual Cortex. *J. Neurosci.* **27**, 10391–10403 (2007).

6. S. LeVay, M. Connolly, J. Houde, D. V. Essen, The complete pattern of ocular dominance stripes in the striate cortex and visual field of the macaque monkey. *J. Neurosci.* **5**, 486–501 (1985).

7. J. C. Horton, D. R. Hocking, An adult-like pattern of ocular dominance columns in striate cortex of newborn monkeys prior to visual experience. *J. Neurosci.* **16**, 1791–1807 (1996).

8. S. Breit, J. B. Schulz, A.-L. Benabid, Deep brain stimulation. *Cell Tissue Res* **318**, 275–288 (2004).

9. J. S. Perlmutter, J. W. Mink, DEEP BRAIN STIMULATION. *Annual Review of Neuroscience* **29**, 229–257 (2006).

10. A. M. Lozano, N. Lipsman, H. Bergman, P. Brown, S. Chabardes, J. W. Chang, K. Matthews, C. C. McIntyre, T. E. Schlaepfer, M. Schulder, Y. Temel, J. Volkmann, J. K. Krauss, Deep brain stimulation: current challenges and future directions. *Nat Rev Neurol* **15**, 148–160 (2019).

11. M. D. Johnson, S. Miocinovic, C. C. McIntyre, J. L. Vitek, Mechanisms and Targets of Deep Brain Stimulation in Movement Disorders. *Neurotherapeutics* **5**, 294–308 (2008).

12. W. M. M. Schüpbach, S. Chabardes, C. Matthies, C. Pollo, F. Steigerwald, L. Timmermann, V. Visser Vandewalle, J. Volkmann, P. R. Schuurman, Directional leads for deep brain stimulation: Opportunities and challenges. *Movement Disorders* **32**, 1371–1375 (2017).

13. S. Zhang, M. Tagliati, N. Pouratian, B. Cheeran, E. Ross, E. Pereira, Steering the Volume of Tissue Activated With a Directional Deep Brain Stimulation Lead in the Globus Pallidus Pars Interna: A Modeling Study With Heterogeneous Tissue Properties. *Front. Comput. Neurosci.* **14** (2020).

14. S. Zhang, P. Silburn, N. Pouratian, B. Cheeran, L. Venkatesan, A. Kent, A. Schnitzler, Comparing Current Steering Technologies for Directional Deep Brain Stimulation Using a Computational Model That Incorporates Heterogeneous Tissue Properties. *Neuromodulation* **23**, 469–477 (2020).

15. G. Nagel, D. Ollig, M. Fuhrmann, S. Kateriya, A. M. Musti, E. Bamberg, P. Hegemann, Channelrhodopsin-1: a light-gated proton channel in green algae. *Science* **296**, 2395–2398 (2002).

16. E. S. Boyden, F. Zhang, E. Bamberg, G. Nagel, K. Deisseroth, Millisecond-timescale, genetically targeted optical control of neural activity. *Nat Neurosci* **8**, 1263–1268 (2005).

17. J. Delbeke, L. Hoffman, K. Mols, D. Braeken, D. Prodanov, And Then There Was Light: Perspectives of Optogenetics for Deep Brain Stimulation and Neuromodulation. *Front Neurosci* **11**, 663 (2017).

18. M. M. Chernov, R. M. Friedman, G. Chen, G. R. Stoner, A. W. Roe, Functionally specific optogenetic modulation in primate visual cortex. *Proceedings of the National Academy of Sciences* **115**, 10505–10510 (2018).

19. O. Ruiz, B. R. Lustig, J. J. Nassi, A. Cetin, J. H. Reynolds, T. D. Albright, E. M. Callaway, G. R. Stoner, A. W. Roe, Optogenetics through windows on the brain in the nonhuman primate. *Journal of Neurophysiology* **110**, 1455–1467 (2013).

20. R. Scharf, T. Tsunematsu, N. McAlinden, M. D. Dawson, S. Sakata, K. Mathieson, Depth-specific optogenetic control in vivo with a scalable, high-density μLED neural probe. *Sci Rep* **6**, 28381 (2016).

21. H. Thair, A. L. Holloway, R. Newport, A. D. Smith, Transcranial Direct Current Stimulation (tDCS): A Beginner's Guide for Design and Implementation. *Front Neurosci* **11**, 641 (2017).

22. N. Bolognini, T. Ro, Transcranial magnetic stimulation: disrupting neural activity to alter and assess brain function. *J Neurosci* **30**, 9647–9650 (2010).

23. R. Beisteiner, M. Hallett, A. M. Lozano, Ultrasound Neuromodulation as a New Brain Therapy. *Advanced Science* **10**, 2205634 (2023).

24. Y. Tufail, A. Yoshihiro, S. Pati, M. M. Li, W. J. Tyler, Ultrasonic neuromodulation by brain stimulation with transcranial ultrasound. *Nat Protoc* **6**, 1453–1470 (2011).

25. G. Li, W. Qiu, Z. Zhang, Q. Jiang, M. Su, R. Cai, Y. Li, F. Cai, Z. Deng, D. Xu, H. Zhang, H. Zheng, Noninvasive Ultrasonic Neuromodulation in Freely Moving Mice. *IEEE Transactions on Biomedical Engineering* **66**, 217–224 (2019).

26. Y. Yang, J. Yuan, R. L. Field, D. Ye, Z. Hu, K. Xu, L. Xu, Y. Gong, Y. Yue, A. V. Kravitz, M. R. Bruchas, J. Cui, J. R. Brestoff, H. Chen, Induction of a torpor-like hypothermic and hypometabolic state in rodents by ultrasound. *Nat Metab* **5**, 789–803 (2023).

27. Z. Lin, L. Meng, J. Zou, W. Zhou, X. Huang, S. Xue, T. Bian, T. Yuan, L. Niu, Y. Guo, H. Zheng, Non-invasive ultrasonic neuromodulation of neuronal excitability for treatment of epilepsy. *Theranostics* **10**, 5514–5526 (2020).

28. F. Munoz, A. Meaney, A. Gross, K. Liu, A. N. Pouliopoulos, D. Liu, E. E. Konofagou, V. P. Ferrera, Long term study of motivational and cognitive effects of low-intensity focused ultrasound neuromodulation in the dorsal striatum of nonhuman primates. *Brain Stimulation* **15**, 360–372 (2022).

29. R. Beisteiner, E. Matt, C. Fan, H. Baldysiak, M. Schönfeld, T. Philippi Novak, A. Amini, T. Aslan, R. Reinecke, J. Lehrner, A. Weber, U. Reime, C. Goldenstedt, E. Marlinghaus, M. Hallett, H. Lohse-Busch, Transcranial Pulse Stimulation with Ultrasound in Alzheimer's Disease—A New Navigated Focal Brain Therapy. *Advanced Science* **7**, 1902583 (2020).

30. S. Jiménez-Gambín, N. Jiménez, J. M. Benlloch, F. Camarena, Generating Bessel beams with broad depth-of-field by using phase-only acoustic holograms. *Sci Rep* **9**, 20104 (2019).

31. G. Antonacci, D. Caprini, G. Ruocco, Demonstration of self-healing and scattering resilience of acoustic Bessel beams. *Applied Physics Letters* **114**, 013502 (2019).

32. X. Zhuang, J. He, J. Wu, X. Ji, Y. Chen, M. Yuan, L. Zeng, A Spatial Multitarget Ultrasound Neuromodulation System Using High-Powered 2-D Array Transducer. *IEEE Transactions on Ultrasonics, Ferroelectrics, and Frequency Control* **69**, 998–1007 (2022).

33. H. Chan, H.-Y. Chang, W.-L. Lin, G.-S. Chen, Large-Volume Focused-Ultrasound Mild Hyperthermia for Improving Blood-Brain Tumor Barrier Permeability Application. *Pharmaceutics* **14**, 2012 (2022).

34. Z. Du, G. Chen, Y. Li, N. Zheng, J.-X. Cheng, C. Yang, Photoacoustic: A Versatile Nongenetic Method for High-Precision Neuromodulation. *Acc. Chem. Res.* **57**, 1595–1607 (2024).

35. Y. Jiang, H. J. Lee, L. Lan, H. Tseng, C. Yang, H.-Y. Man, X. Han, J.-X. Cheng, Optoacoustic brain stimulation at submillimeter spatial precision. *Nat Commun* **11**, 881 (2020).

36. L. Shi, Y. Jiang, Y. Zhang, L. Lan, Y. Huang, J.-X. Cheng, C. Yang, A fiber optoacoustic emitter with controlled ultrasound frequency for cell membrane sonoporation at submillimeter spatial resolution. *Photoacoustics* **20**, 100208 (2020).

37. L. Shi, Y. Jiang, F. R. Fernandez, G. Chen, L. Lan, H.-Y. Man, J. A. White, J.-X. Cheng, C. Yang, Non-genetic photoacoustic stimulation of single neurons by a tapered fiber optoacoustic emitter. *Light Sci Appl* **10**, 143 (2021).

38. Y. Li, Y. Jiang, L. Lan, X. Ge, R. Cheng, Y. Zhan, G. Chen, L. Shi, R. Wang, N. Zheng, C. Yang, J.-X. Cheng, Optically-generated focused ultrasound for noninvasive brain stimulation with ultrahigh precision. *Light Sci Appl* **11**, 321 (2022).

39. H. Estrada, J. Robin, A. Özbek, Z. Chen, A. Marowsky, Q. Zhou, D. Beck, B. le Roy, M. Arand, S. Shoham, D. Razansky, High-resolution fluorescence-guided transcranial ultrasound mapping in the live mouse brain. *Science Advances* **7**, eabi5464 (2021).

40. H. Hocheng, C.-M. Chen, Y.-C. Chou, C.-H. Lin, Study of novel electrical routing and integrated packaging on bio-compatible flexible substrates. *Microsyst Technol* **16**, 423–430 (2010).

41. B. G. Sanganahalli, G. J. Thompson, M. Parent, J. V. Verhagen, H. Blumenfeld, P. Herman, F. Hyder, Thalamic activations in rat brain by fMRI during tactile (forepaw, whisker) and non-tactile (visual, olfactory) sensory stimulations. *PLOS ONE* **17**, e0267916 (2022).

42. F. A. Duck, Medical and non-medical protection standards for ultrasound and infrasound. *Progress in Biophysics and Molecular Biology* **93**, 176–191 (2007).

43. Y. Choe, J. W. Kim, K. K. Shung, E. S. Kim, Microparticle trapping in an ultrasonic Bessel beam. *Applied Physics Letters* **99**, 233704 (2011).

44. Z. Xu, W. Xu, M. Qian, Q. Cheng, X. Liu, A flat acoustic lens to generate a Bessel-like beam. *Ultrasonics* **80**, 66–71 (2017).

45. N. Jiménez, R. Picó, V. Sánchez-Morcillo, V. Romero-García, L. M. García-Raffi, K. Staliunas, Formation of high-order acoustic Bessel beams by spiral diffraction gratings. *Phys. Rev. E* **94**, 053004 (2016).

46. J.-Y. Lu, J. F. Greenleaf, Ultrasonic nondiffracting transducer for medical imaging. *IEEE Transactions on Ultrasonics, Ferroelectrics, and Frequency Control* **37**, 438–447 (1990).

47. W. Gao, W. Liu, Y. Hu, J. Wang, Study of Ultrasonic Near-Field Region in Ultrasonic Liquid-Level Monitoring System. *Micromachines (Basel)* **11**, 763 (2020).

48. J. Durnin, J. J. Miceli, J. H. Eberly, Comparison of Bessel and Gaussian beams. *Opt. Lett., OL* **13**, 79–80 (1988).

49. J. K. Brynildsen, L.-M. Hsu, T. J. Ross, E. A. Stein, Y. Yang, H. Lu, Physiological characterization of a robust survival rodent fMRI method. *Magnetic Resonance Imaging* **35**, 54–60 (2017).

50. P. A. Valdes Hernandez, A. Sumiyoshi, H. Nonaka, R. Haga, E. Aubert Vasquez, T. Ogawa, Y. Iturria Medina, J. J. Riera, R. Kawashima, An in vivo MRI Template Set for Morphometry, Tissue Segmentation, and fMRI Localization in Rats. *Front. Neuroinform.* **5** (2011).

**Acknowledgments**

We acknowledge Runyu Wang and Michael Marar for their assistance in OBUS device fabrication and in vivo experiment setup. Immunofluorescence imaging reported in this publication was supported by the Boston University Micro and Nano Imaging Facility and the Office of the Director, National Institutes of Health of the National Institutes of Health under award Number S10OD024993. The content is solely the responsibility of the authors and does not necessarily represent the official views of the National Institute of Health.

**Funding:**
Focused Ultrasound Foundation grant (CY). NIH R21 EY035437-01 (CY) supporting G.C. R01NS109794 (JXC).

**Author contributions:**
    Conceptualization: YL, LL
    Methodology: YL, TR, NT
    Investigation: YL, GC, TR, NT, YZ Z, CM, NZ
    Supervision: CY, JX C, NM
    Writing—original draft: YL, TR
    Writing—review & editing: YL, TR, CY, JX C

**Competing interests:** C.Y. and J.X.C. serve as Scientific Advisor for Axorus. CY received a research grant from Axorus, which did not support this work. C.Y. and J.X.C. have a patent on Methods and Devices for Optoacoustic Stimulation (US Patent No. 11684404 B2) issued. Other authors claim no COI.

**Data and materials availability:** The raw fMRI data that support the findings of this study are available from the corresponding author upon reasonable request. All other data are available in the main text or the supplementary materials.

# Supplementary Materials for

**Miniaturized optically-generated Bessel beam ultrasound for volumetric transcranial brain stimulation**

Yueming Li et al.

*Corresponding author. Email: cheyang@bu.edu (C.Y.); jxcheng@bu.edu (JX.C.).

**This PDF file includes:**

    Figs. S1 to S14
    Tables S1

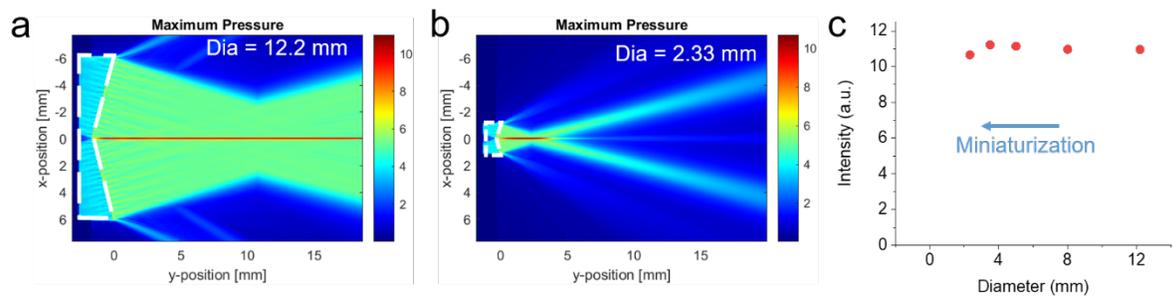

**Fig. S1. Acoustic field intensity is maintained for miniaturized OBUS.** The white dashed line indicates the location of the acoustic emitters.

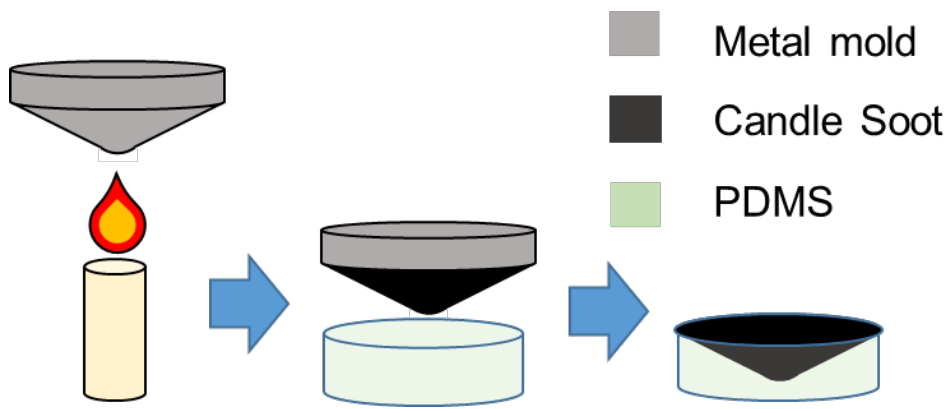

**Fig. S2. Fabrication process of the OBUS.**

Candle soot was deposited onto the metal mold through flame synthesis and subsequently transferred into PDMS, followed by the curing process.

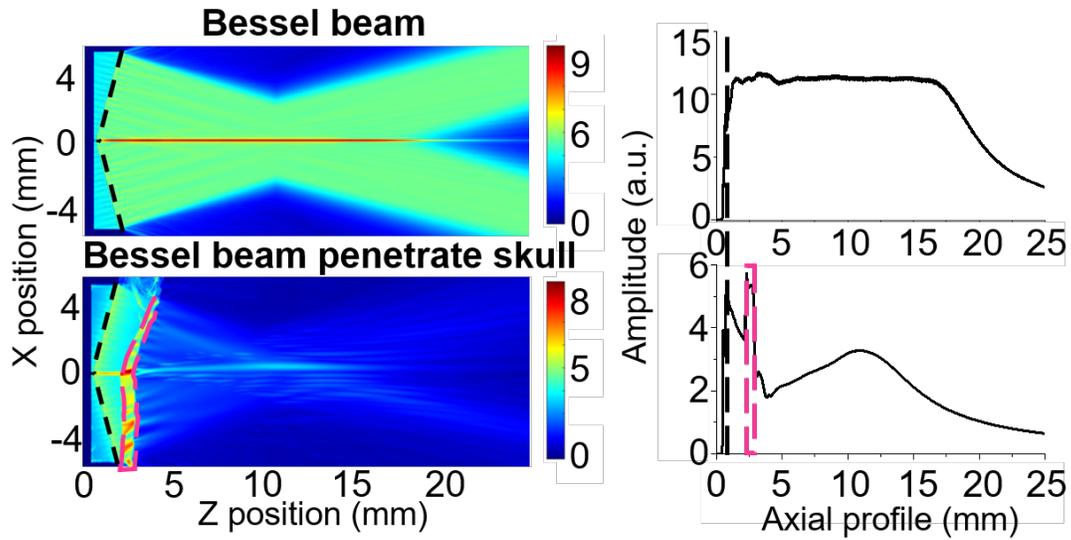

**Fig. S3. The simulation presents the acoustic field of a standard acoustic Bessel beam.**

A. Acoustic field of a standard acoustic Bessel beam before and after interacting with the rat skull. Black dot line: the surface of the emitter. Pink region: the position of the rat skull. B. The profile of the maximum pressure point along the axial direction. Black dot line: the surface of the emitter. Pink rectangular: the position of the rat skull.

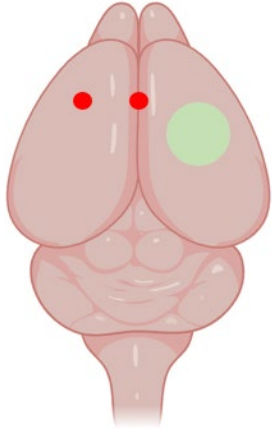

**Fig. S4. Schematic showing the locations of punctures (red dots, Bregma and untreated side) in horizontally sliced brain tissue relative to the targeted area (green circle).**

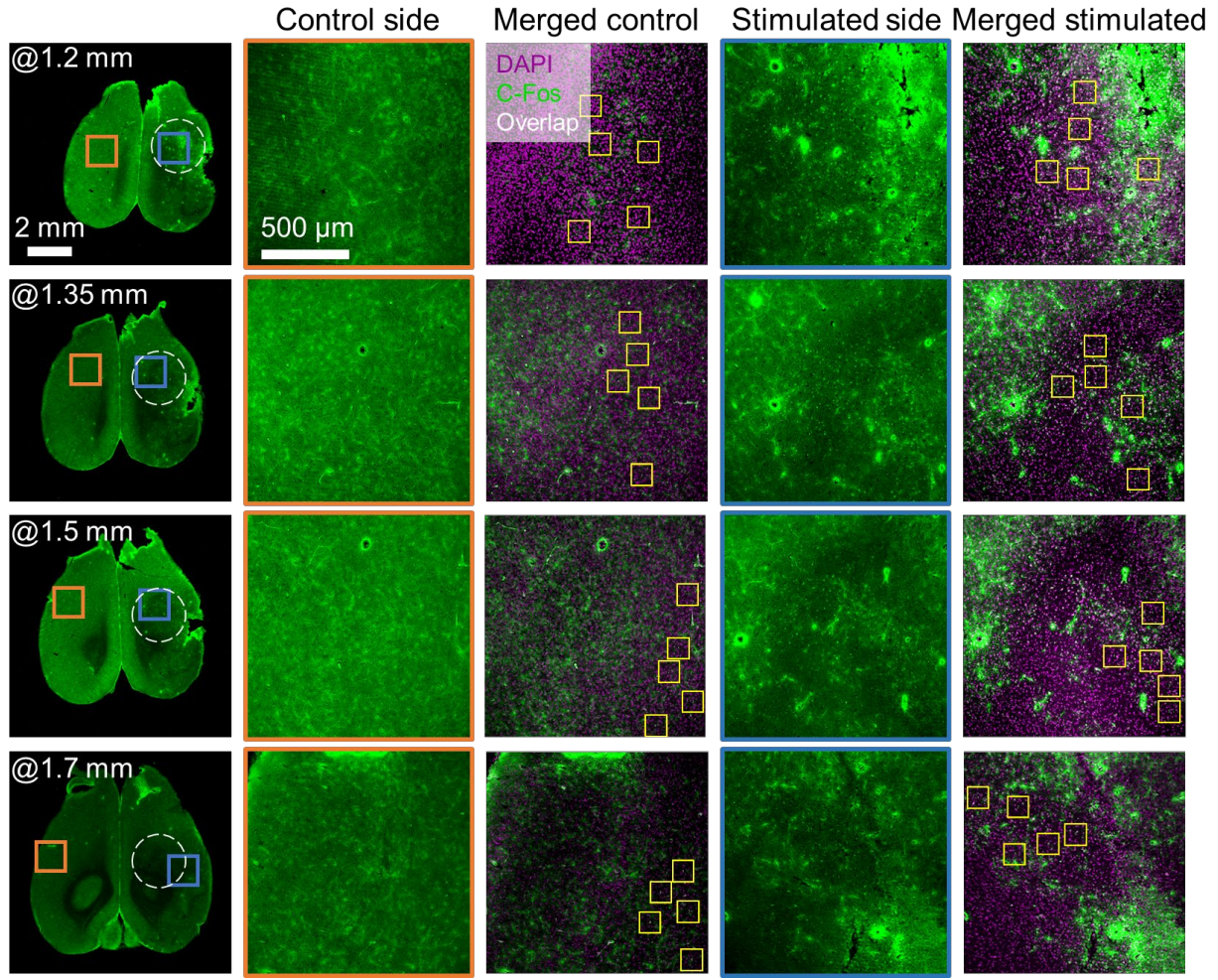

**Fig. S5. Immunofluorescence examination of c-fos expression at depths ranging from 1.2 mm to 1.7 mm in the mouse brain.**

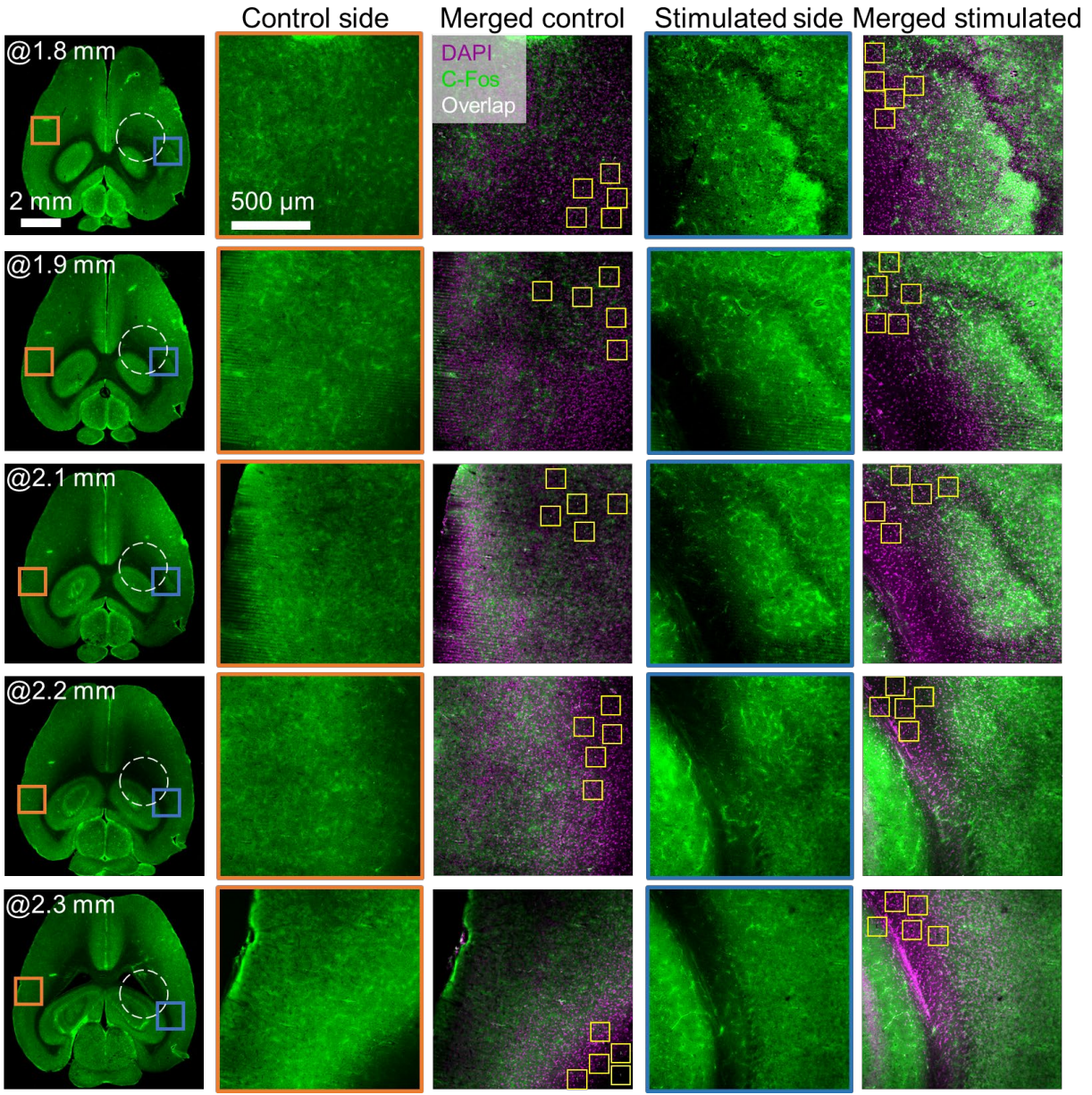

**Fig. S6. Immunofluorescence examination of c-fos expression at depths ranging from 1.8 mm to 2.3 mm in the mouse brain.**

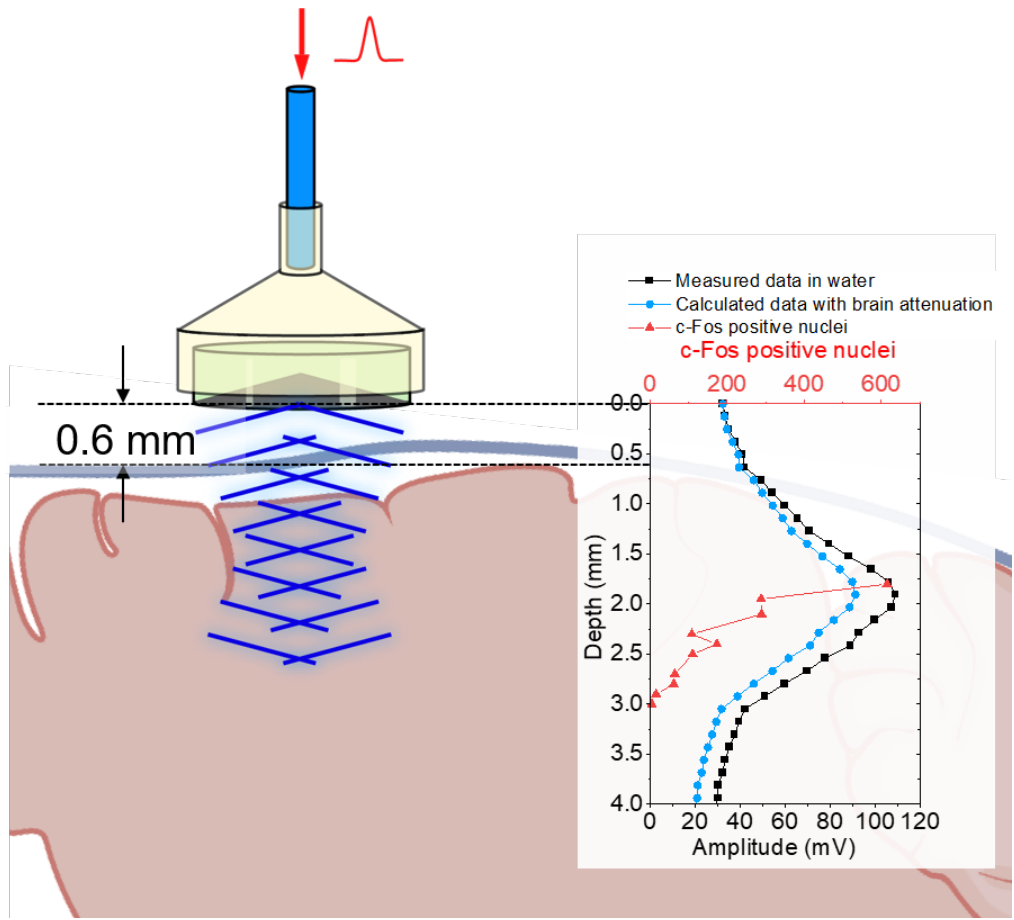

**Fig. S7. The decay pattern of the c-Fos signal with the axial acoustic profile.**

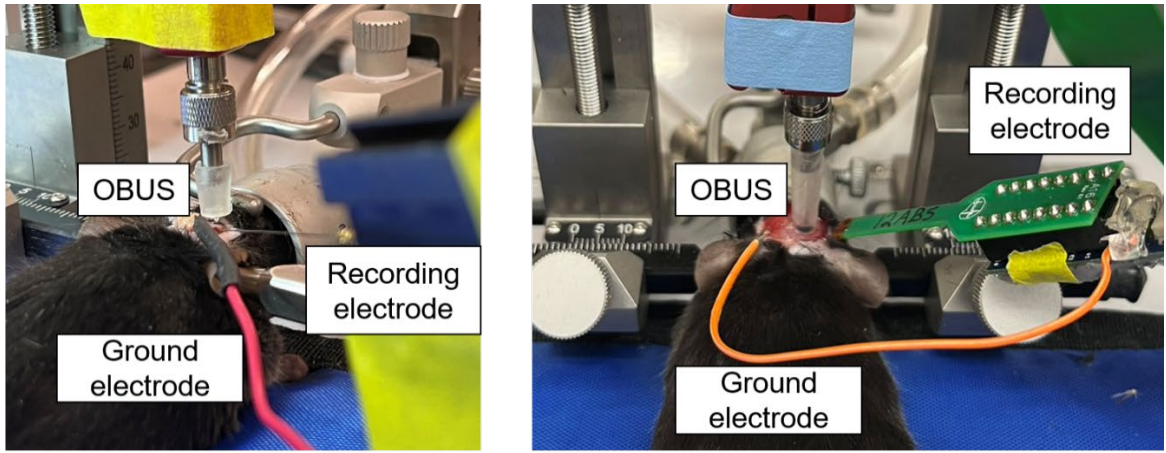

**Fig. S8. A.** A photo illustrating the alignment process using a 3D printer pointer to align the VTA of OBUS with the tip of the recording electrode. **B.** A photo of the recording setup for OBUS in vivo stimulation.

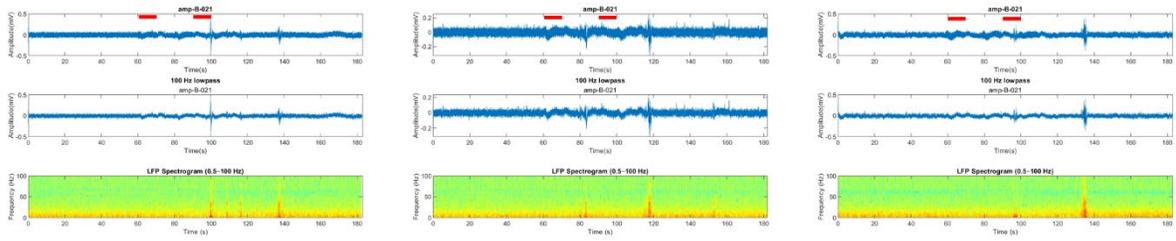

**Fig. S9. Raw data, 100 Hz low pass filtered data, and corresponding spectrogram of the three repeats when applying OBUS with 2.8 MPa.**

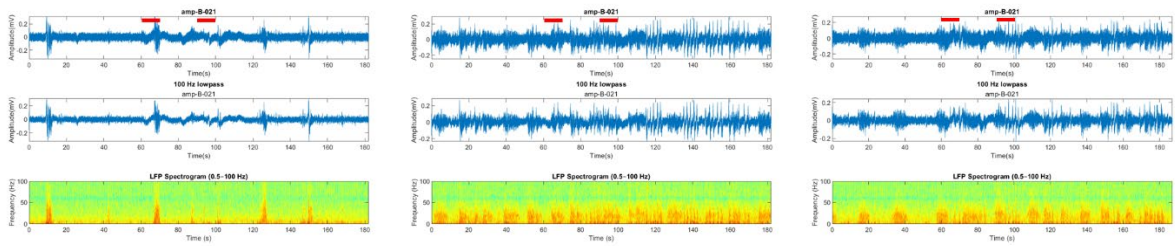

**Fig. S10. Raw data, 100 Hz low pass filtered data, and corresponding spectrogram of the three repeats when applying OBUS with 3.7 MPa.**

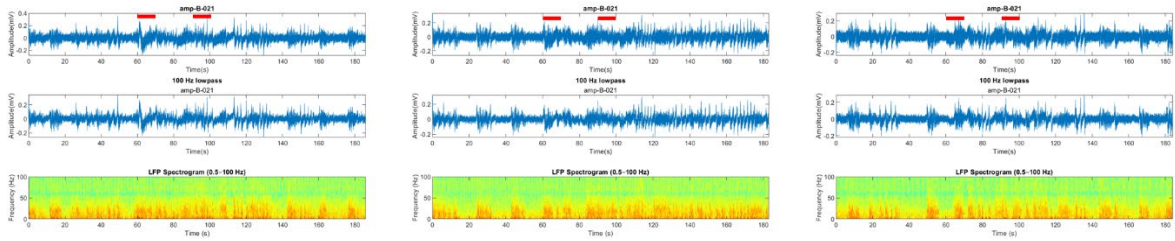

**Fig. S11. Raw data, 100 Hz low pass filtered data, and corresponding spectrogram of the three repeats when applying OBUS with 4.1 MPa.**

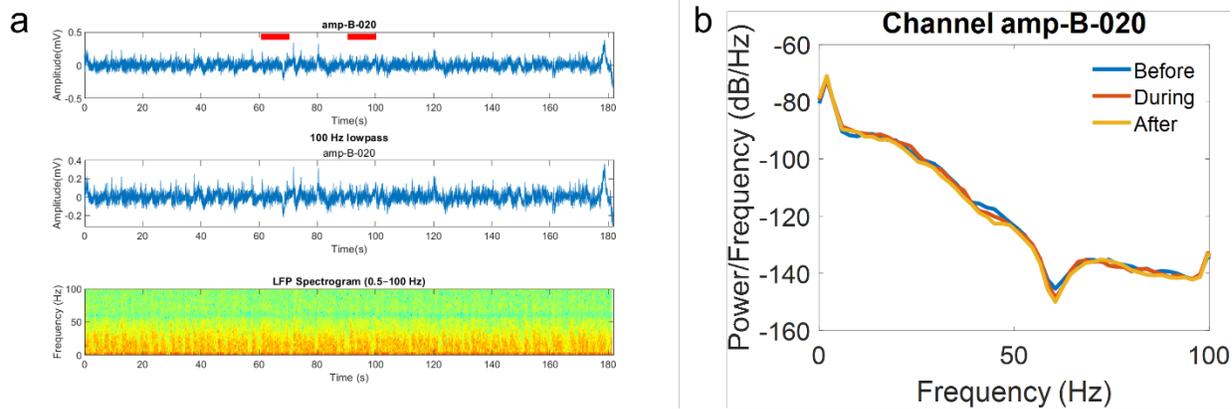

**Fig. S12.** A. Raw trace, 100 Hz lowpass and corresponding spectrogram of the 0 mW laser output control, with red lines indicating the onset of the laser. B: PSD calculated for this control.

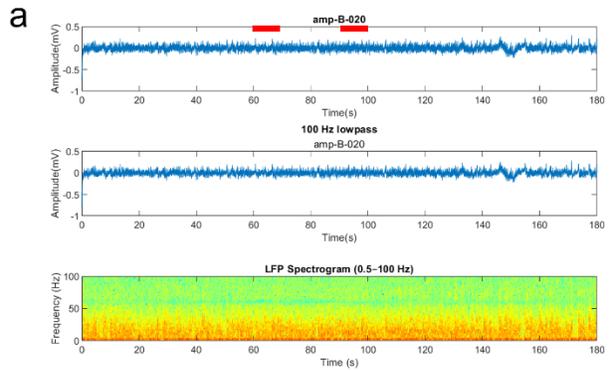 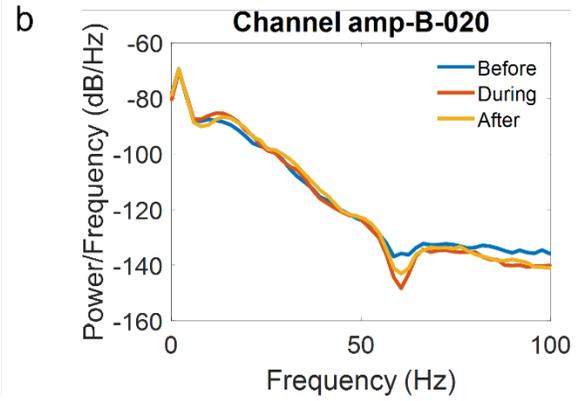

**Supplementary Figure 13 A. Raw trace, 100 Hz lowpass and corresponding spectrogram of the 3.8 MPa targeting the opposite side of S1, with red lines indicating the onset of the laser. B: PSD calculated for this control.**

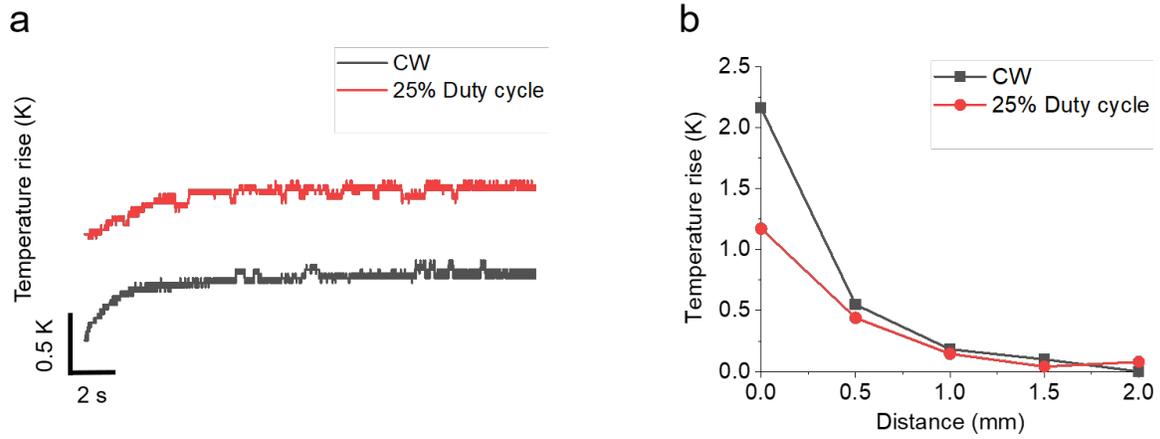

**Fig. S14. Temperature below the OBUS under CW and stimulation conditions, measured by a thermalcoupler.**
A. The temperature profiles measured at 0.5 mm away from the OBUS surface. B. The temperature rise of the OBUS measured at different depths in various modes.

|  | Standard Bessel Beam |
|---|---|
| Peak intensity transcranial efficiency | 28.6% |
| Axial resolution (mm) – pre-skull penetration | 19.88 |
| Axial resolution (mm) – post-skull penetration | 13.65 |
| Percentage of change in axial resolution | 68.7% |
| Lateral resolution (mm) – pre-skull penetration | 0.38 |
| Lateral resolution (mm) – post-skull penetration | 0.98 |
| Percentage of change in lateral resolution | 260.0% |

**Table S1. Summary of peak intensity transcranial efficiency and axial and lateral resolution changes pre- and post- skull penetration for standard Bessel Beam.**